\documentclass[twocolumn]{aastex62}
\usepackage{natbib}
\usepackage{color} 
\usepackage{aas_macros} 
\usepackage{hyperref} 

\received{2022}
\revised{}
\accepted{}

\begin{document}

\title[Galactic Mass with halo stars using Jeans and TME]{Milky Way Mass with K Giants and BHB Stars Using LAMOST,
  SDSS/SEGUE, and {\it Gaia}: 3D Spherical Jeans Equation and Tracer
  Mass Estimator}

\email{xuexx@nao.cas.cn, sarahbird@ctgu.edu.cn, liuchao@nao.cas.cn,} 
\email{cflynn@swin.edu.au, jtshen@sjtu.edu.cn} 

\author{Sarah A. Bird} 
\affil{College of Science, China Three Gorges University, Yichang 443002, People's Republic of China} \affil{Center for Astronomy and Space Sciences, China Three Gorges University, Yichang 443002, People's Republic of China}
\affil{CAS Key Laboratory of Optical Astronomy, National Astronomical Observatories, Chinese Academy of Sciences, Beijing 100101, People's Republic of China}
\affil{Shanghai Astronomical Observatory, Chinese Academy of Sciences, 80 Nandan Road Shanghai 200030, People's Republic of China}

\author{Xiang-Xiang Xue}
\affil{CAS Key Laboratory of Optical Astronomy, National Astronomical Observatories, Chinese Academy of Sciences, Beijing 100101, People's Republic of China}
\affil{School of Astronomy and Space Science, University of Chinese Academy of Sciences, Beijing 100049, People's Republic of China}

\author{Chao Liu}
\affil{Key Laboratory of Space Astronomy and Technology, National Astronomical Observatories, Chinese Academy of Sciences, Beijing 100101, People's Republic of China}
\affil{School of Astronomy and Space Science, University of Chinese Academy of Sciences, Beijing 100049, People's Republic of China}

\author{Chris Flynn} \affil{Centre for Astrophysics and Supercomputing, Swinburne
  University of Technology, Post Office Box 218, Hawthorn, VIC 3122,
  Australia}

\author{Juntai Shen} 
\affil{Department of Astronomy, School of Physics and Astronomy, Shanghai Jiao Tong University, 800 Dongchuan Road, Shanghai 200240, People's Republic of China}
\affil{Key Laboratory for Particle Astrophysics and Cosmology (MOE) / Shanghai Key Laboratory for Particle Physics and Cosmology, Shanghai 200240, People's Republic of China}
\affil{Shanghai Astronomical Observatory, Chinese Academy of Sciences, 80 Nandan Road Shanghai 200030, People's Republic of China}

\author{Jie Wang}
\affil{Key Laboratory for Computational Astrophysics, National Astronomical Observatories, Chinese Academy of Sciences, 20A Datun Road, Beijing 100101, People's Republic of China}
\affil{School of Astronomy and Space Science, University of Chinese Academy of Sciences, Beijing 100049, People's Republic of China}
  
\author{Chengqun Yang}
\affil{Shanghai Astronomical Observatory, Chinese Academy of Sciences, 80 Nandan Road Shanghai 200030, People's Republic of China}

\author{Meng Zhai}
\affil{CAS Key Laboratory of Optical Astronomy, National Astronomical Observatories, Chinese Academy of Sciences, Beijing 100101, People's Republic of China}
\affil{Chinese Academy of Sciences South America Center for Astronomy, National Astronomical Observatories, Chinese Academy of Sciences, Beijing 100012, People's Republic of China}

\author{Ling Zhu}
\affil{Shanghai Astronomical Observatory, Chinese Academy of Sciences, 80 Nandan Road Shanghai 200030, People's Republic of China}

\author{Gang Zhao}
\affil{CAS Key Laboratory of Optical Astronomy, National Astronomical Observatories, Chinese Academy of Sciences, Beijing 100101, People's Republic of China}
\affil{School of Astronomy and Space Science, University of Chinese Academy of Sciences, Beijing 100049, People's Republic of China}

\author{Hai-Jun Tian} \affil{College of Science, China Three Gorges University, Yichang 443002, People's Republic of China} \affil{Center for Astronomy and Space Sciences, China Three Gorges University, Yichang 443002, People's Republic of China}

\begin{abstract}

We measure the enclosed Milky Way mass profile to Galactocentric distances of $\sim70$ and $\sim50$ kpc using the smooth, diffuse stellar halo samples of Bird et al. The samples are LAMOST and SDSS/SEGUE K giants (KG) and SDSS/SEGUE blue horizontal branch (BHB) stars with accurate metallicities. The 3D kinematics are available through LAMOST and SDSS/SEGUE distances and radial velocities and {\it Gaia} DR2 proper motions. Two methods are used to estimate the enclosed mass: 3D spherical Jeans equation and Evans et al. tracer mass estimator (TME). We remove substructure via the Xue et al. method based on integrals of motion. We evaluate the uncertainties on our estimates due to random sampling noise, systematic distance errors, the adopted density profile, and non-virialization and non-spherical effects of the halo. The tracer density profile remains a limiting systematic in our mass estimates, although within these limits we find reasonable agreement across the different samples and the methods applied. Out to $\sim70$ and $\sim50$ kpc, the Jeans method yields total enclosed masses of $4.3\pm0.95$ (random) $\pm0.6$ (systematic) $\times10^{11}$ M$_\odot$ and $4.1\pm1.2$ (random) $\pm0.6$ (systematic) $\times10^{11}$ M$_\odot$ for the KG and BHB stars, respectively. For the KG and BHB samples we find a dark matter virial mass of $M_{200}=0.55^{+0.15}_{-0.11}$ (random) $\pm0.083$ (systematic) $\times10^{12}$ M$_\odot$ and $M_{200}=1.00^{+0.67}_{-0.33}$ (random) $\pm0.15$ (systematic) $\times10^{12}$ M$_\odot$, respectively.

\end{abstract}


\section{Introduction} \label{sec:intro}

The Milky Way is typical of large spiral galaxies, in possessing a
bright disk and bulge embedded in a tenuous, approximately spherical
stellar halo, comprised of stars, globular clusters, satellite
galaxies, and a dark matter halo which dominates its mass budget.

The kinematic properties of these halo objects can be used to probe
the enclosed mass as a function of Galactocentric radius $r$.
Historically, satellite galaxies and globular clusters have allowed us
to probe the mass distribution to large distances, beyond 100 kpc from
the Galactic center. In an influential study, \citet{Battaglia2005}
used 240 halo objects including stars, globular clusters and satellite
galaxies to map the mass profile to $\approx 120$ kpc, finding an
enclosed mass of $0.8^{+1.2}_{-0.2} \times 10^{12}$ M$_\odot$.

It is clear that large samples of objects are needed to reduce
sampling noise and reveal any kinematic substructures which must be
removed or otherwise accounted for in the analysis. In the last few
years, the total sample size of halo stars with line-of-sight
velocities, distances, and stellar parameters has reached of order
$10^4$, due to the substantial efforts made in several dedicated
facilities to obtain large numbers of medium to high resolution
stellar spectra using fiber-fed spectrographs, such as the Large Sky
Area Multi-Object Fiber Spectroscopic Telescope (LAMOST) survey
\citep{Cui2012,Deng2012,Luo2012,Zhao2012}, Sloan Digital Sky Survey
(SDSS) / SEGUE \citep{Yanny2009}, {\it Gaia}-ESO Survey
\citep{Gilmore2012,Randich2013}, RAVE \citep{Steinmetz2006}, and Galah
\citep{Martell2017}.  A range of automated techniques have been
developed to find appropriate tracers, including K-giant (KG) stars
\citep{Kafle2014}, blue horizontal branch (BHB) stars
\citep{Xue2008,Kafle2012}, RR Lyrae stars \citep{Cohen2017}, reaching
in some cases beyond 100 kpc from the Galactic center.

The Milky Way's enclosed mass is thus being probed with increasingly
larger samples of individual stellar halo tracers. \citet{Xue2008}
used 2401 SDSS/SEGUE halo BHB stars; \citet{Gnedin2010} used 910
Hypervelocity Star Survey halo BHB stars and blue stragglers, mapping the
mass profile out to 80 kpc; \citet{Kafle2012} used 4664 SDSS/SEGUE
halo BHB stars to constrain the profile to $\approx 60$ kpc and extended
the work in \citet{Kafle2014}, using SDSS/SEGUE halo K-giant and BHB
stars, with additional constraints provided by the gas terminal
velocity curve and Sgr A$^{*}$ proper motion, to measure the virial
mass and virial radius of the Milky Way; and \citet{Huang2016} used
$\approx 16,000$ outer disk red clump giant stars and $\approx 5700$
SDSS/SEGUE halo K giants to measure the mass profile out to $\approx
100$ kpc. More recently, \citet{Zhai2018} have measured the mass
profile to 120 kpc using some 9000 halo K giants.

In most of these studies, only radial velocities have been available
for the tracers, and information about the important transverse
motions of the tracers need to be gleaned statistically (see, e.g.,
\citet{Kafle2012}, \citet{Kafle2014}) from the samples. At large
Galactocentric distances, this becomes increasingly difficult to
perform from radial velocities alone. The European Space Agency's
{\it Gaia} mission \citep{GaiaCollaborationPrusti2016} is providing the proper motions necessary to measure
transverse motions directly; this leads to much improved understanding
of the halo tracer kinematics.

Now with the recent availability of high quality full 6D phase-space information for large numbers of sources, much effort has been made to decrease the uncertainties in the Milky Way mass estimate. Recent works using a tracer mass estimator with 6D phase-space information include \citet[][globular clusters]{Sohn2018}, \citet[][globular clusters]{Watkins2019}, and \citet[][satellites]{Fritz2020}. The most recent work using the spherical Jeans equation by \citet{Zhai2018} is very similar to our current investigation in method and data (LAMOST K giants) although only line of sight velocities were included, whereas we additionally make use of proper motions from {\it Gaia} to obtain the stellar tangential velocities. Using Bayesian analysis to fit a distribution function to full 6D phase-space data (globular clusters, satellites, halo stars) has been a recent popular choice among many works \citep{Posti2019,Eadie2019,Vasiliev2019,Deason2021,Correa_Magnus2022,Shen2022,Slizewski2022,Wang_Jianling2022} and a similar distribution function analysis using 5D phase-space data from {\it Gaia} \citep{Hattori2021}.
In addition to fitting the observational data with a distribution function, several works have incorporated into the fitting a comparison of the observed data with Milky Way-type galaxies from cosmological simulations \citep{Callingham2019,Li_Zhao-Zhou2020}. Newly discovered high velocity stars with full 6D phase-space information have been used to estimate the mass of the Milky Way \citep{Monari2018,Hattori2018.866,Deason2019.485,Grand2019,Koppelman2021,Necib2022b}. \citet{Vasiliev2021} and \citet{Craig2022} have estimated the Milky Way mass by fitting models for the Sagittarius and Magellanic Steams, respectively. Several recent studies have estimated the Milky Way mass using measurements of the rotation curve \citep{Eilers2019,de_Salas2019,Ablimit2020,Karukes2020,Cautun2020,Jiao2021}. Other works have used 6D satellite phenomenology, characterizing simulated Milky Way-type satellite populations and comparing to the observations of satellites in the Milky Way, to estimate the mass of the Milky Way \citep{Patel2018,Rodriguez_Wimberly2022,Villanueva-Domingo2022}. \citet{Zaritsky2020} apply the timing argument to distant Milky Way halo stars to derive a lower limit to the Milky Way mass.

In this article, we apply the Jeans theorem \citep{Jeans1915}, which relates the density and kinematics
of particles moving within a gravitational potential, assuming they
are in a steady state. The simplest form is to assume spherical
symmetry for the potential, the stellar tracer density, and the tracer
velocity dispersion profiles. It has been used to measure the Milky
Way gravitational potential \citep[e.g.,][]{Battaglia2005, Dehnen2006,
  Xue2008, Gnedin2010, Samurovic2011, Kafle2012, Bhattacharjee2014,
  Huang2016, Ablimit2017, Zhai2018} using line-of-sight velocities,
and making assumptions about the transverse velocities of the tracer
stars (i.e., their anisotropy).  \citet{Wang_Wenting2020} have recently
reviewed Milky Way mass estimates of this type.

{\it Gaia} proper motions for tracer stars now make possible the use of
the Jeans equation on the full 3D position and velocity data, rather
than line-of-sight velocities alone.

In this paper, we use the 3D velocity dispersion profiles of
\citet{Bird2019beta,Bird2021} and, for the first time, use the 3D
spherical Jeans equation to measure the Milky Way mass profile. The
paper is laid out as follows. In Section \ref{sec:meth} we introduce
the two methods which we use for Galactic mass estimation, the 3D
spherical Jeans equation and the tracer mass estimator (TME) of
\citet{Evans2011}, both of which assume the simplest case halo
dynamics, that of a spherical system traced by a non-rotating, relaxed
population in equilibrium.
The use of two mass estimate methods allows for comparison
to tie the
new determination back to previous estimates of the Milky Way's
enclosed mass.
After presenting the methods in Section \ref{sec:meth}, in Section
\ref{sec:uncertainties} we analyze the sources of random and
systematic uncertainty on our mass estimates. 
These are
effects due to the Gaia-Enceladus-Sausage (GES) satellite and Large Magellanic Cloud (LMC),
residual substructure in the halo sample, and non-spherical effects.
In Section \ref{sec:dataresults}, we apply the Jeans and TME methods to our large
sample of halo K giants and BHB stars.  
We extrapolate the mass determination to make an
estimate of the Milky Way's virial mass, virial radius, and
concentration parameter, assuming the dark halo has an NFW profile. 
We discuss and summarize our results and conclusions in Section
\ref{sec:discussion}.

\section{Methods} \label{sec:meth}

We apply two methods to our data to recover the Milky Way's enclosed
mass profile, firstly, the spherical Jeans equation using full 3D
distances and velocities, and secondly a tracer mass estimator (TME)
method, which uses the distances, tracer Galactocentric radial
velocities, and anisotropy parameter measured from the 3D velocities.

The density fall-off of the tracer stars is modeled as a power-law
$\rho\propto r^{\alpha}$. This assumption remains one of the important
sources of uncertainty in our final mass estimates. We discuss this
assumption in detail in Sections \ref{sec:density-prof} and \ref{subsec:dist-and-den}.

In order to build the mass profile using our two methods, 
we bin our stars according to Galactocentric radius, selecting separate radial bins for our KG and BHB samples. We elaborate upon our bin selection later in Section \ref{sec:bins}. Once selected, we use the same binning scheme for both the 3D spherical Jeans method and TME.

We also extrapolate our results to estimate the virial mass and radius
of the Milky Way, assuming an NFW dark matter profile.

In this section
we describe the details of applying each method and estimating the associated uncertainties propagated from the observed data.

\subsection{3D Spherical Jeans Equation} \label{sec:3DJeans}

We use the 3D spherical Jeans equation for a non-rotating, stationary
system of stars, to measure the Galactic mass profile $M(<r)$, where
$M$ is the mass enclosed within the Galactocentric radius $r$:

\begin{equation} \label{eq:mass}
M(<r) = -\frac{1}{G}(r^2\frac{d\sigma_r^2}{dr} + r((2+\alpha)\sigma_r^2 - \sigma_\theta^2 - \sigma_\phi^2)).
\end{equation}

Here, the gravitational constant is $G=0.054( (4\pi) \times 10^{-3} )$
(km s$^{-1}$)$^2$ kpc $M_\odot^{-1}$, and the velocity dispersions in
the radial and two transverse directions are given by
$(\sigma_r,\sigma_\theta, \sigma_\phi)$, and $\alpha$ is the slope
of the power-law fit to the density fall-off of the tracers $\rho\propto r^{\alpha}$. In Section \ref{sec:density-prof} we discuss $\alpha$ and
the density profiles used for the modeling in detail, and in Section \ref{sec:bins} we detail our selected radial bins which we use to calculate the mass profile.

The 3D velocity dispersion profiles are computed from our sample
following \citet{Bird2021}. We use the 3D version of the software {\tt extreme-deconvolution} \citep{Bovy2011} which takes into account
the individual velocity errors on each star in each
component $(r,\theta,\phi)$, the covariances between the velocity errors, and
the velocity means. This software implements a maximum likelihood density estimation technique in order to model the underlying, error-deconvolved distribution as a mixture of Gaussian components.

The Jeans equation has a term which is the gradient of the radial
component of the velocity dispersion. We fit $\sigma_r(r)$ with a
smooth, continuously differentiable function, in order to estimate
this term. We find that an exponential form for $\sigma_r(r)$ fits the data well and is bound and well-behaved at the ends of the profile. The choice of an exponential function was determined by trial-and-error in which we also tested linear and spline functions. We select the model as a fitting function only, without direct physical interpretation.

Thus, for $\sigma_r$ we have,

\begin{equation} \label{eq:sigr}
\sigma_r=\sigma_0\exp{\left[-\frac{r}{h_r}\right]},
  \label{eqn:s}
\end{equation}

and the differential form:

\begin{equation} \label{eq:sigr2}
\frac{d\sigma_r^2}{dr}=-\frac{2\sigma_0^2}{h_r}\exp{\left[-\frac{2r}{h_r}\right]}.
  \label{eqn:s2}
\end{equation}

We perform a Bayesian Markov chain Monte Carlo (MCMC) maximum likelihood method and make use of the {\tt
  PYTHON} package {\tt emcee} \citep{emcee}. We optimize Equation
\ref{eq:sigr} and find the best fit for $\sigma_0$ and $h_r$. We use
1000 walkers, 1500 runs, and 500 runs for the burn-in phase. We select minimal priors, 
requiring $0<h_r<1000$ kpc. We use the 50th percentile as the best fit for ($\sigma_0$, $h_r$) and we investigate the goodness-of-fit using the 16th and 84th percentiles as the random uncertainty.

We note that, although we parameterize the $\sigma_r$ profile in order to measure the $\frac{d\sigma_r^2}{dr}$ term needed for the Jeans equation, in our application of the Jeans equation we use a hybrid of binned and parameterized velocity data, i.e., we use the binned velocity dispersion measurements from {\tt extreme-deconvolution} and the parameterized $\frac{d\sigma_r^2}{dr}$ from our Bayesian MCMC estimate.

Observational errors in each term of the Jeans equation are propagated through to derive uncertainties in the enclosed mass estimate as follows. Firstly, to estimate the error on the velocity dispersions in radial bins, we use the Poisson statistics of the number of stars in each bin (after using {\tt extreme-deconvolution} to determine the velocity dispersion, taking into account the individual observational errors on the stellar velocities, and the covariances between them). Secondly, to estimate the error on the the gradient of the radial velocity dispersion term in the Jeans equation, we propagate the errors on the radial velocity dispersions, also determined using Poisson statistics of the star numbers in the radial bins. Monte Carlo simulations of the terms in the Jeans equation are then performed with 100 realizations using Gaussian distributed errors as above, to derive a final uncertainty on the enclosed mass profile.

\subsection{Tracer Mass Estimator (TME)}
\label{sec:tme}

Estimation of the mass of a gravitationally bound system typically
involves calculating a weighted average of the velocities and
distances. Traditionally, mass estimators have been based on the
virial theorem with pioneering work by \citet[][]{Zwicky1933,
  Chandrasekhar1942, Schwarzschild1954, Burbidge1959}; and
\citet{Limber1960}. Virial mass estimators take the form of two
separate averages for distance and velocity. More convenient forms of
the mass estimator, involving the average over the combination of
distance and velocity, have been introduced by \citet{Holmberg1937,
  Page1952, Hartwick1978}; and \citet{Bahcall1981}. These authors took differing
approaches to derive their mass estimators, including consideration of
the orbital properties of the system, use of the collisionless
Boltzmann equation and/or the Jeans equation.

The projected mass estimator was introduced by
\citet{Bahcall1981}, who noted that the virial mass estimator has
limitations, being both biased and inefficient (see Section II of
\citet{Bahcall1981} and also \citet{An2011}). The application is to
data for which the line-of-sight velocities at projected distances
from the center of the mass distribution are available, so as is the
case for galaxy clusters.

We use a tracer mass estimator of \citet{Evans2003,Evans2011}. They are the generalization of the projected mass
estimator to the case in which the mass (or number) density of the
tracer population differs from the overall mass density, and distances
and velocities for the tracers are available. The \citet{Evans2011}
tracer mass estimator is particularly convenient for this study, as it
makes direct use of our observables (distances, line-of-sight
velocities, proper motions), and has been tested on simulated data
\citep[such as galaxies formed in $N$-body simulations,][]{Evans2003,Evans2011}.

As with the Jeans equation, the TME assumes a spherical system of
non-rotating relaxed tracers which is in dynamical equilibrium. This holds to a good approximation in our data set except for bulk motion in the outer halo due to the Large Magellanic Cloud.

We use the scale-free tracer mass estimator of \citet[Equation 9]{Evans2011}\footnote{For a full derivation of the estimator, see \citet{An2011}.}. As we will elaborate upon in Section \ref{section:lmc}, we incorporate a correction for the effects of the Large Magellanic Cloud by subtracting the mean radial velocity $\langle V_r\rangle$ from our radial velocities $V_{r,i}$. Thus the equation becomes,
\begin{equation}
  M(<r)\approx\frac{r^{0.5}(\gamma-\alpha-2\beta)}{GN}\sum^{N}_{i=1}r_i^{0.5}(V_{r,i}-\langle V_r\rangle)^2.
\label{eqn:mass}
\end{equation}
Here, we have $N$ tracer stars where the $i$th tracer has
Galactocentric distance $r_i$ and radial velocity $V_{r,i}$ to yield
the mass $M(<r)$ of the Galaxy within the radius $r$ of the most
distant star in the sample. To build the $M(<r)$ mass profile, we apply Equation \ref{eqn:mass} to our KG and BHB stars in bins of Galactocentric radius, using the same binning scheme as we use for our application of the 3D spherical Jeans method. In this manner, $\gamma$, $\alpha$, and $\beta$ are functions of $r$. The Galactic potential is represented as a
power-law with slope $\gamma$ as a function of $r$. The tracer stars
are assumed to have a density fall-off as a power-law with slope
$\alpha$
(as adopted for the Jeans equation\footnote{Our use of the symbols $\gamma$ and $\alpha$ differs to other
  works (e.g., \citet{Watkins2010}, \citet{Evans2011}, \citet{An2011}) in order to avoid confusion
  with our adopted form of the Jeans equation.}). 
The velocity anisotropy profile along Galactocentric radius
of the tracer stars is given to be $\beta$, where
\begin{equation}
  \beta = 1 -  (\sigma_\theta^2 + \sigma_\phi^2 ) / (2\sigma_r^2).
  \label{eqn:beta}
\end{equation}
To measure $\beta$,
we calculate the 3D velocity dispersion profiles, as in our application of the Jeans equation, by following \citet{Bird2021} using the {\tt extreme-deconvolution} software \citep{Bovy2011}. Allowing $\beta$ to vary along $r$ is appropriate as has been previously shown, e.g., in analyses of simulated Milky Way-type galaxies \citep{Loebman2018,Cunningham2019b} and in the Milky Way itself \citep{Bird2019beta,Cunningham2019b,Bird2021}.

In Section \ref{sec:density-prof} we discuss $\alpha$ and
the density profiles used for the modeling in detail, and in Section \ref{sec:bins} we detail our selected radial bins which we use to calculate the mass profile.

In an analysis of NFW potentials covering a wide range of virial radii and concentrations,
\citet{Watkins2010}
found that a scale-free host potential with a power-law index of
$\gamma=0.5$ well represents the NFW potential for large galaxies like the
Milky Way. We adopt $\gamma=0.5$ to analyze our sample, since many
studies have shown the success of using an NFW model to approximate
dark halo mass distributions.

We estimate the random uncertainty on the TME mass by bootstrapping the stellar sample. For each radial bin, we bootstrap 100 new samples, calculate the TME mass using the newly sampled ($r$, $V_r$), and take the quadrature sum of the 16th and 84th percentiles as the random uncertainty on the mass.

\subsection{Estimation of the Virial Mass}
\label{sec:method_m200}

The Jeans and TME methods yield the enclosed mass of the Milky Way,
which represents the sum of baryonic and dark components. In order
to estimate the mass of the dark component, we fit the enclosed
mass profiles after subtracting a model of the radial distribution of
the Milky Way's baryons. 

Specifically, we fit the Jeans and TME enclosed mass profiles (Sections
\ref{sec:jeansmass} and \ref{sec:tmemass}, respectively) with a model
of the Milky Way baryons \citep[consisting of a bulge and disk component][]{Bovy2013,Bovy2015}
and an NFW dark matter profile.

The bulge and disk models are taken from the {\tt MWPotential2014}
model within the {\tt galpy.potential} module of the {\tt PYTHON}
library for Galactic dynamics {\tt galpy} \citep{Bovy2015}.  We adopt
the values $R_\odot=8.3$ kpc and
$v_\mathrm{circ}(r_\odot)=220$ km s$^{-1}$ for the distance of the Sun
from the Galactic center and the solar circular velocity
\citep{Bovy2012,Bovy2015}.

The bulge component is modeled as a power-law density profile that is
exponentially cut-off \\ {\tt (PowerSphericalPotentialwCutoff)}. We
adopt a power-law exponent of $\alpha_\mathrm{bulge} = -1.8$, mass
$M_b = 0.5 \times 10^{10}$ M$_\odot$, and cut-off radius of
$r_\mathrm{bulge} = 1.9$ kpc:

\begin{equation}
\rho(r)=\frac{G M_b}{r^{\alpha_\mathrm{bulge}}}\exp(-(r/r_\mathrm{bulge})^2).
\end{equation}

The disk component is modeled as a \citet{MiyamotoNagai1975} disk
({\tt MiyamotoNagaiPotential}) with a disk scale length $a = 3$ kpc,
disk scale height $b = 0.28$ kpc, and mass $M_d = 6.8 \times
10^{10}$ M$_\odot$:

\begin{equation}
\Phi(R,z)=-\frac{G M_d}{\sqrt{R^2+(a+\sqrt{z^2+b^2})^2}}.
\end{equation}

Compared to the total radial force at $R_\odot$, the relative force
contribution from the bulge is 0.05 and from the disk is 0.6. These
relative amplitudes have been fit to the data in \citet{Bovy2015}.

The NFW dark matter component as a function of $r$, as presented in
\citet{Navarro1995.275.720,Navarro1996,Navarro1997} and
\citet[p. 71]{Binney2008}, takes the form

\begin{equation}
M_\mathrm{NFW}(r)=4\pi \rho_0 r_\mathrm{s}^3 \left( \ln(1+r/r_\mathrm{s})-\\
  \frac{r/r_\mathrm{s}}{1+(r/r_\mathrm{s})}     \right),
\label{eqn:nfwmass}
\end{equation}

where

\begin{equation}
\rho_0=\rho_\mathrm{crit}\delta_\mathrm{c},
\end{equation}

\begin{equation}
\delta_\mathrm{c}=\frac{200}{3}\frac{c^3}{[\ln(1+c)-c/(1+c)]},
\end{equation}

\begin{equation}
\rho_\mathrm{crit}=3H^2/(8\pi G)/1000^3,
\end{equation}

\begin{equation}
  r_\mathrm{s}=r_\mathrm{200}/c.
\label{eqn:rs}
\end{equation}

Here, $r_{200}$ is the distance at which the mean density of the
Galaxy is 200 times the present critical density $\rho_\mathrm{crit}$ of the
Universe.

We adopt a Hubble constant of $H=67.8$ km s$^{-1}$ Mpc$^{-1}$
\citep{PlanckXIII2016}. The uncertainty in this parameter affects our
virial mass estimates negligibly compared to other sources of error.

The concentration $c$ and characteristic overdensity
$\delta_\mathrm{c}$ are dimensionless parameters and $r_\mathrm{s}$ is
the NFW scale radius. The concentration and scale radius set the value of
$\rho_0$.  The virial mass $M_{200}$ is the dark mass enclosed within the
spherical volume of radius $r_{200}$,
\begin{equation}
M_{200}=\frac{4\pi}{3}200\rho_\mathrm{crit}r_{200}^3.
\end{equation}

We set the virial radius $r_{200}$ and concentration parameter $c$ of
the NFW profile as free parameters in our fitting. We perform a Bayesian MCMC method and make use of the {\tt
  PYTHON} package {\tt emcee} \citep{emcee}. We optimize Equation
\ref{eqn:nfwmass} and find the best fit for $r_{200}$ and $c$. We use
500 walkers, 400 runs, and 250 runs for the burn-in phase. We select minimal priors, 
requiring $c>0$ and
$200<v_\mathrm{circ}(R_\odot)<250$ km s$^{-1}$. For discussion and comparison, we use the software package {\tt numpy.average} to calculate the average mass and uncertainties of our mass estimates together with other virial mass $M_\mathrm{vir}$ estimates from the literature. These averages are weighted by the errors quoted for each measurement.

\section{Itemizing Sources of Uncertainty} \label{sec:uncertainties}

In the past, measuring the mass profile of the Milky Way has been
strongly affected by random and systematic errors, particularly due to
small sample sizes.

The advent of large automated radial velocity and proper motion
surveys have brought sampling errors down substantially, and also
allowed us to probe other important sources of error, such as the
validity of the well-mixed assumption in applying kinematic methods to
measuring mass.
The assumption that the halo is roughly spherical and in
a stationary dynamical state, even with the careful removal of
substructure, will be incorrect at some level.

In this section we itemize such sources of error on our analysis.
We
first consider the uncertainties involved in our mass estimation and
in the following Section \ref{sec:dataresults} we present our results. 
In Section \ref{sec:GES} we discuss the effects of the ancient merger $Gaia$-Enceladus-Sausage, in Section \ref{section:lmc} we detail our correction for the effects of the LMC, and in Section \ref{subsection:nonvir} we explain our adopted 15 percent systematic uncertainty on our mass estimate due to non-spherical and non-virial effects. These are in addition our estimation of the random sampling uncertainty which we have detailed previously in Section \ref{sec:meth}.

\subsection{Effects of the {\it Gaia}-Enceladus-Sausage Satellite (GES)}
\label{sec:GES}

The structure known as the {\it Gaia}-Enceladus-Sausage \citep[GES,][]{Belokurov2018,Deason2018.862,Haywood2018,Helmi2018,Koppelman2018}
permeates the inner population of halo stars. As this large
satellite merged at an early time in the Galaxy's evolution, most of
its stars have likely virialized and cannot be recognized as
originating in the GES satellite from kinematics alone. Since we clean
substructure from our sample to get at the smooth, virialized
background halo, the GES will remain as part of the sample.

On the other hand, \citet{Lancaster2019} are able to adequately model
the velocity ellipsoid of BHB halo stars using two Gaussian
components, one representing the GES and one an `isotropic stellar
halo.' As we use our smooth, diffuse halo star sample from which we
have removed the streams and substructure via the stellar orbital
parameters \citep[][in preparation]{Xue2019}, we use a single Gaussian model for the
velocity ellipsoid to describe our sample.  Further work on fitting
the model of \citet{Lancaster2019} to our smooth, diffuse halo sample
is presented in \citet{Wu2022}. If the smooth, diffuse halo is a
composition of radial GES stars and isotropic stars combined from
other mergers and events, it is possible measuring the density profile
for each component and using these in the 3D Jeans mass estimation may
further improve the mass estimate. This would be significantly more
complex modeling than what we use in this paper, concentrating on the
diffuse, smooth component only.

\subsection{Effects of the Large Magellanic Cloud (LMC)}
\label{section:lmc}

\citet{Petersen2021} and \citet{Erkal2021} have used distant halo
stars to measure the present-day reflex (or sloshing) motion of the
stellar halo which is induced by the LMC. The LMC's mass induces
systematic motions in the outer halo, affecting the measured
velocities in Galactic longitude $V_l$ (and in the vertical direction,
$V_Z$), with the largest influence occurring for $r>50$ kpc.

Sloshing of the halo by the LMC has been examined using simulations
\citep{Gomez2015, Erkal2020b, Garavito-Camargo2019, Cunningham2020,Petersen2020, Wang_Jianling2022}. Specifically, \citet{Erkal2020b}
run Milky Way models to quantify the bias introduced in mass determinations of the Milky Way
for this effect, since the assumption of equilibrium dynamics is
broken. This bias can cause an overestimation of the Milky Way mass by 25 percent.
They use the TME as presented by \citet{Watkins2010} and find that, by incorporating a correction to the velocity term by subtracting the mean velocity, the bias is greatly reduced. They find that the bias changes depending on the observed quadrant of the sky. In several regions, the bias is negligible after applying the correction to the mass estimate.

\citet{Deason2021} apply their distribution function mass measurement method to
the Milky Way models of \citet{Erkal2020b} and estimate the correction to the velocities needed for their method.
They then apply their method to a sample of distant halo stars (their distant halo K-giant and BHB
stars ($r>50$ kpc) and our halo star samples are taken from the
same catalogs).
They note that the sample stars (fortuitously) occupy a region
of configuration space in the halo which is least affected by the LMC,
so for the sample stars here, the correction is small. Failing to account
for the LMC's sloshing results in overestimating the Milky Way enclosed mass
by $\approx 6-7$ percent for the region of the sky in our halo star sample.

Evidence in support of a massive LMC on its first infall \citep[e.g.,][]{Besla2007,Kallivayalil2013,Laporte2018} was one of the main motivations driving \citet{Slizewski2022} to revisit the distribution function mass estimation method of \citet{Eadie2019}.
\citet{Slizewski2022} measure the Galactic mass using dwarf galaxies separated into different distance intervals and, as investigated by \citet{Erkal2020b}, different spatial octants.
The largest difference they find is the mass measured from stars within the range of $30<r<45$ kpc. They conjecture a possible cause is the influence of the LMC on the velocities of these stars such as caused by the sloshing \citep{Petersen2021,Erkal2021} or the wake \citep{Conroy2021}.

Following the recommended velocity correction of \citet{Erkal2020b},
we take the mean velocity into account for both the velocity
dispersions (using \\{\tt extreme-deconvolution}) of the 3D spherical
Jeans and the velocity component in the \citet{Evans2011} TME used in
this work. Applying the mean velocity corrections, we find that the
changes in the mass profile are very small (few percent) compared to
other sources of error. We include the mean velocity when calculating
the enclosed mass profiles.

\subsection{Non-Virialization, Non-Spherical, and Non-NFWness Effects}
\label{subsection:nonvir}

Mass estimators generally assume that the dark matter and/or stellar
halo tracer is kinematically virialized and stationary (i.e., has a
steady state distribution function) and that the mass distribution is
spherical. Frequently the dark matter halo is assumed to follow an NFW profile. Biases in mass estimates occur when these assumptions do not adequately characterize the galaxy and stellar halo being studied \citep[e.g.,][]{Yencho2006,Wang2015,Han2016b,Sanderson2017,Wang2017,Eadie2018,Wang2018,Grand2019,Deason2021}.

Studies find that non-virialization and non-spherical effects tend towards underestimating the mass when using stellar tracers \citep[e.g.][]{Yencho2006,Wang2018,Grand2019,Deason2021,Rehemtulla2022}. Recently \citet{Deason2021} apply the distribution function method to stellar halos from Auriga \citep{Grand2017} and find that the mass is typically
underestimated by $\sim10$ percent, with a scatter of $\sim25$ percent.

\citet{Wang2017} analyze the extent to which these
assumptions affect the mass estimates of dark halos in cosmological
simulations. They find that spherical, NFW mass profile fits to the dark matter
halos can typically induce errors in the mass estimates of 25 percent.

The NFW concentration $c$ parameter (Equation \ref{eqn:rs}) depends on the scale radius which can be from one to a few $\times 10$ kpc for Milky Way-type galaxies. Thus, the inner halo data are needed to constrain $c$, as noted by, e.g., \citet{Kafle2014} and \cite{Wang2017}. 

On the other hand, the addition of baryons to simulations can cause the inner halo to deviate from an NFW profile \citep[e.g.,][]{Schaller2015,Dutton2016,Lovell2018} which can warrant for excluding the innermost halo samples. In such cases, \citet{Wang2017} find that
the use of stellar halo tracers can
yield systematically underestimated galaxy masses and overestimated concentration. The bias is highest when they use stars with $r>1$ kpc but progressively decreases as they exclude inner halo stars, e.g., by keeping stars with $r>10$ kpc or $r>20$ kpc to measure the mass.

In this study, we adopt an error in our mass estimates due to non-spherical and non-virialization effects of 15 percent 
following \citet{Deason2021}. In attempts to alleviate possible deviations from the NFW profile in the inner halo, we subtract the baryon mass profile from our measured total mass profile before fitting with an NFW model; additionally, we require the circular velocity near the Sun to be $200<v_\mathrm{circ}<250$ km s$^{-1}$. We further discuss the effects of the inner halo stars on the mass estimate in Section \ref{subsec:dist-and-den}.

\section{Data Analysis and Results} \label{sec:dataresults}

\subsection{Data: Two Smooth, Diffuse Stellar Halo Samples} \label{sec:data}

\begin{figure}
\includegraphics[width=.95\columnwidth]{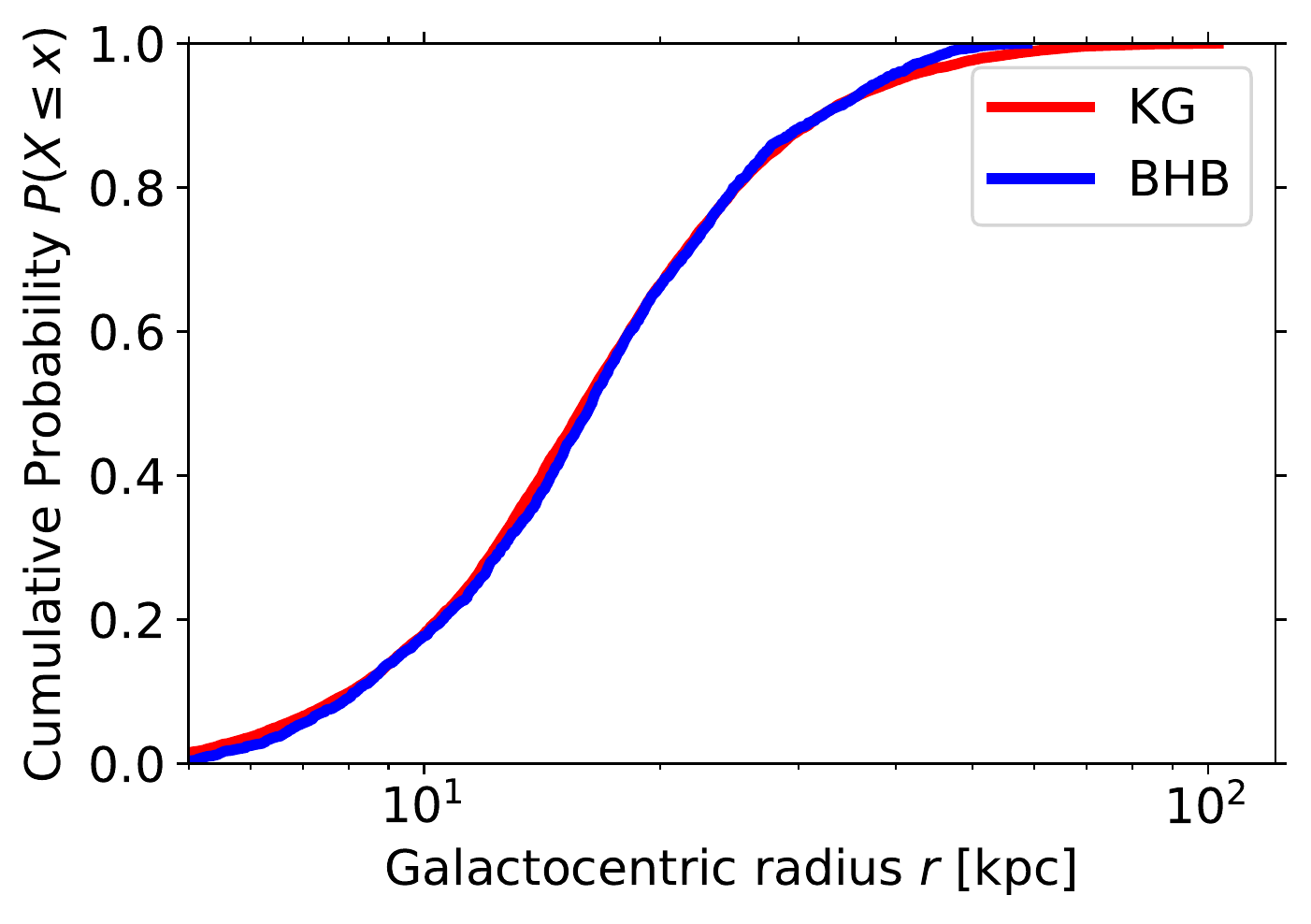}
\includegraphics[width=.95\columnwidth]{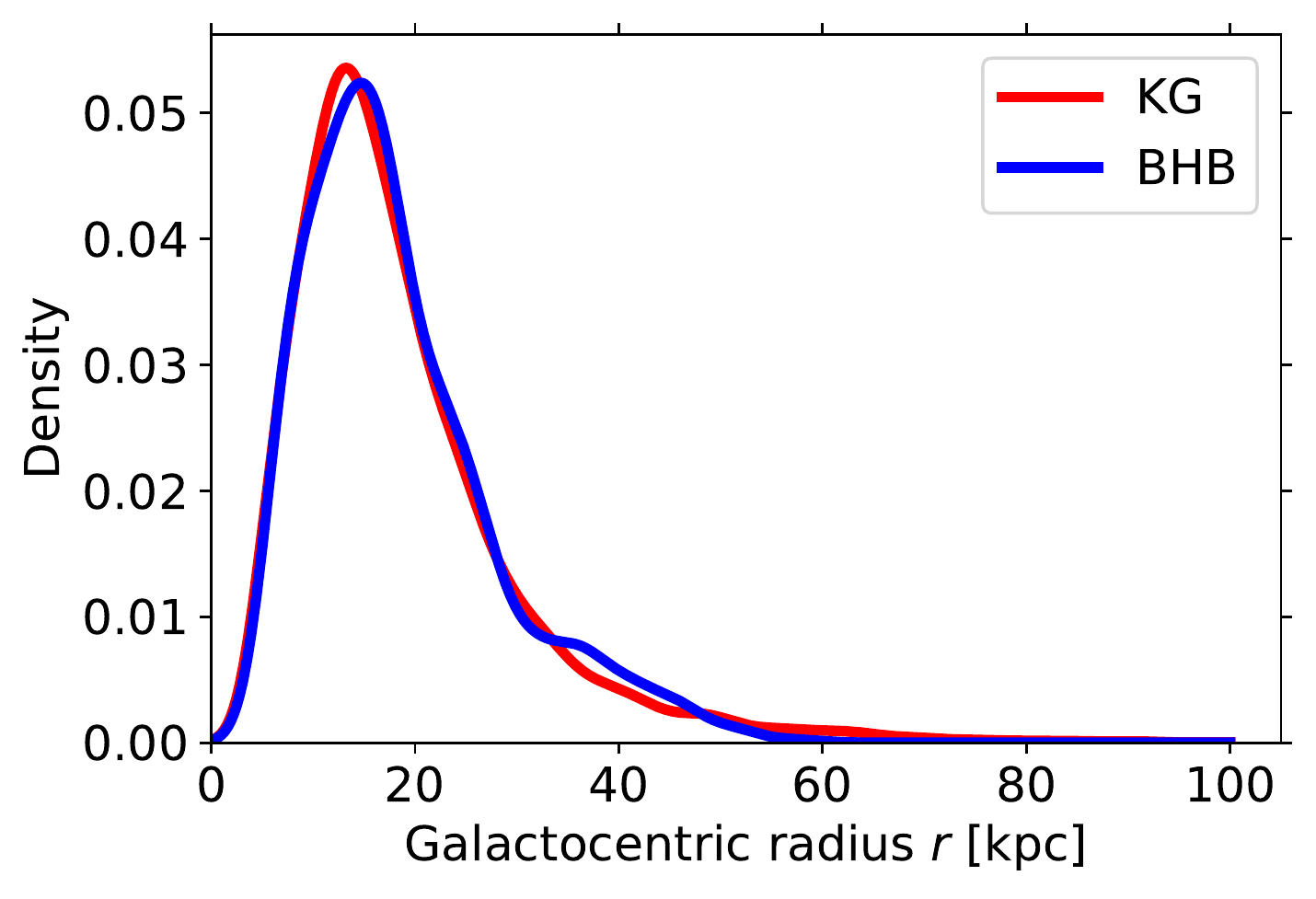}
\includegraphics[width=.95\columnwidth]{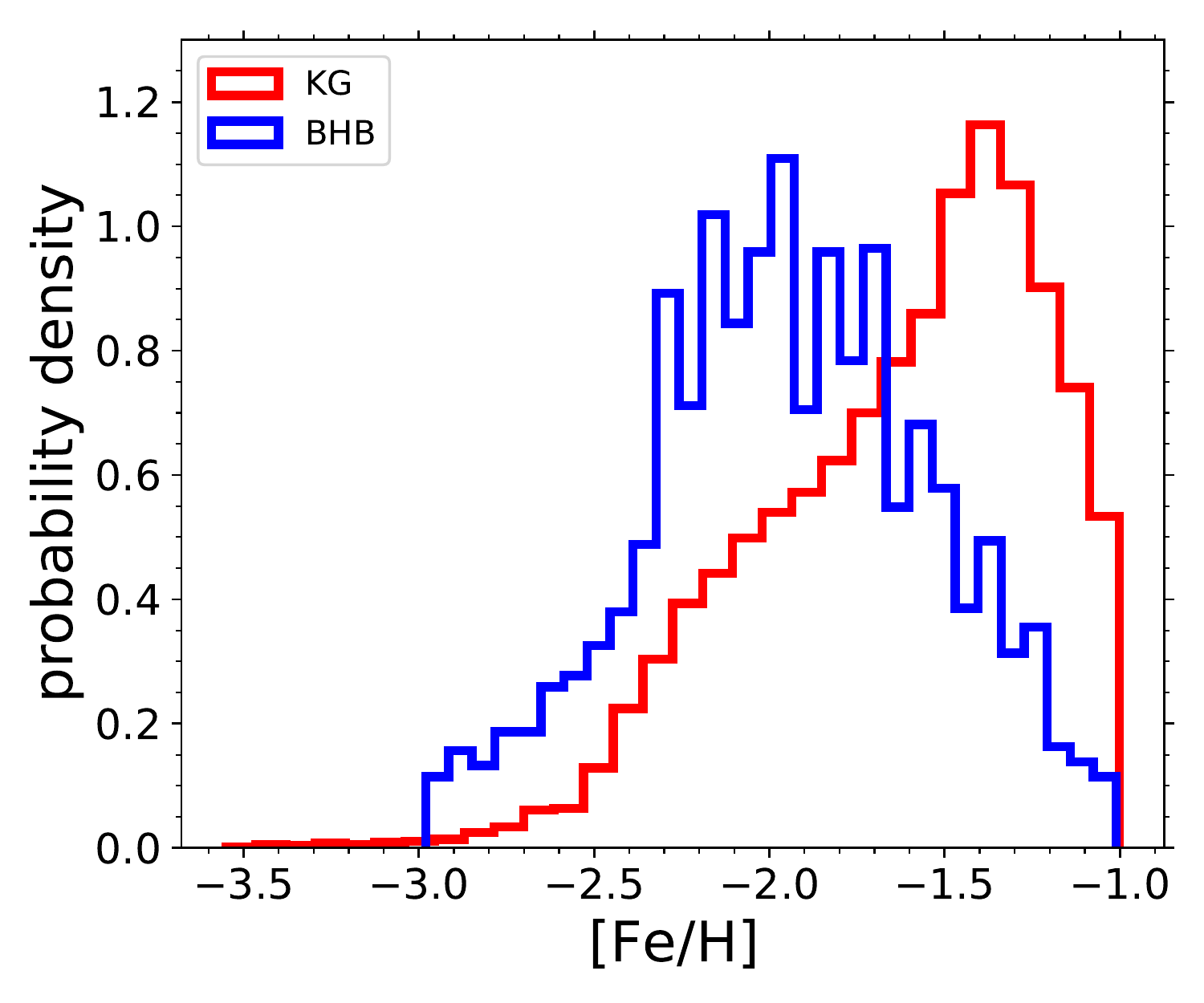}
\caption{Cumulative probability (upper panel), one dimensional kernel density estimation (middle panel), and probability density distribution of [Fe/H] (lower panel) using KG (red) and BHB (blue) smooth, diffuse halo stars. The $x$-axis of the upper panel is in log scale and of the middle panel linear.
  More than 50 percent of our smooth halo KG and BHB samples lie within 25 kpc; this is within the break radius where the fall in the density profile changes as discussed in Section \ref{sec:density-prof} and shown in Figure \ref{fig:density}. Our KG sample peaks at higher metallicities compared to our BHB sample.
} 
  \label{fig:r}
\end{figure}

\begin{figure*}
\includegraphics[width=2\columnwidth]{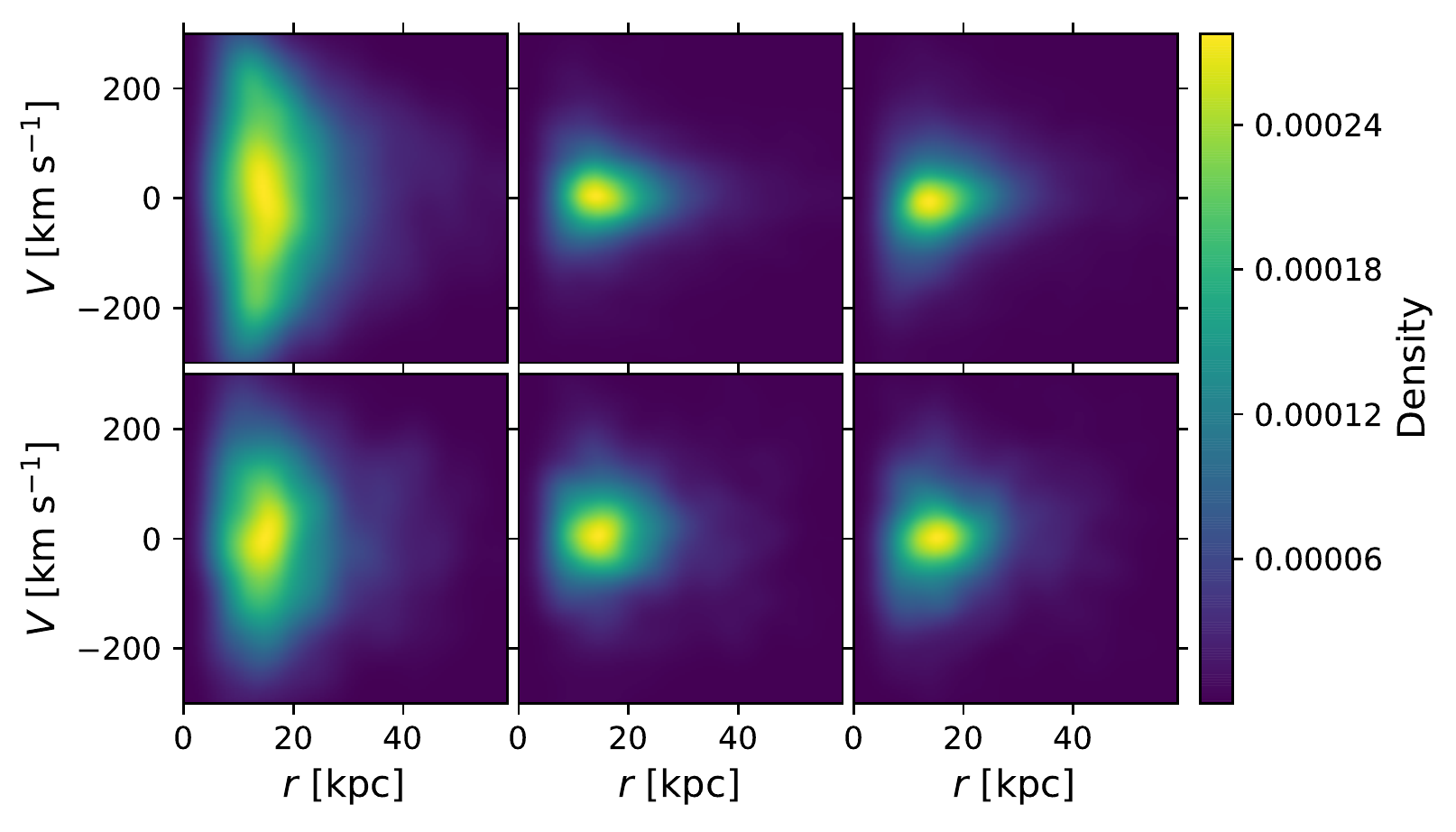}
\caption{Two dimensional kernel density estimation for the 3D spherical Galactocentric velocity components (from left to right, $V_r$, $V_\theta$, and $V_\phi$) and Galactocentric distance $r$ using KG (upper rows) and BHB (lower rows) stars. 
} 
  \label{fig:V}
\end{figure*}

We use the LAMOST K-giant, SDSS K-giant, and SDSS BHB halo star
samples of \citet{Bird2021}. These consist of line-of-sight
velocities and metallicities obtained with LAMOST and SDSS/SEGUE, star
distance estimates using the method of \citet{Xue2014} and
\citet{Xue2008} for K giants and BHB stars (respectively), and finally
proper motions from {\it Gaia} DR2 \citep{GaiaCollaborationBrown2018}. The LAMOST/SDSS KG and BHB samples initially
contain 12728, 5248, and 3982 stars \citep[Table 2][]{Bird2021}, respectively.

Our LAMOST KG line-of-sight velocities have been corrected by adding 8 km s$^{-1}$ to account for the systematic bias found by comparing stars in common with our SDSS/SEGUE KG sample \citep{Bird2021}. This bias has been previously detected and discussed in the literature although its origin remains unknown \citep{Luo2015,Tian2015,Xiang2015,Schonrich2017,Yang2019b,Wang_Wenting2020}.

The Galactocentric Cartesian Coordinates follow the conventions in \citet{astropy2018} and Galactocentric spherical coordinates follow \citet{Bird2019beta} and \citet{Bird2021}. The Sun is located at
$X_\odot=-8.3$ kpc and $Z_\odot=29$ pc. The local standard of rest
(LSR) velocity is $v_\mathrm{LSR}=220$ km s$^{-1}$ , and the motion of the Sun
with respect to the LSR is $(U, V, W)_\odot=(11.1, 12.2,7.3)$ km s$^{-1}$ \citep{Schonrich2010}. These are similar to the more recent measurements of the LSR parameters \citep{Huang2015,Tian2015}.

As expressed in \citet{Bird2021}, our samples include the propagation of (1) uncertainties in the distances
of the stars, (2) line-of-sight radial velocity
uncertainties, 
(3) two {\it Gaia} DR2
proper motion error estimates, and (4) the {\it Gaia} DR2 proper motion covariances. These four uncertainties are propagated via 1000 Monte Carlo runs per star with the {\tt python} package {\tt pyia} \citep{Price-Whelan2018} to derive the uncertainties on the 3D kinematics of each star in the Galactocentric Cartesian and spherical coordinate systems.

Both the 3D spherical Jeans equation and TME assume the population
tracing the Galactic potential is a relaxed and virialized system;
we, thus, use the smooth, diffuse halo samples from which streams and
substructure have been removed using the method of \citet[in
  preparation]{Xue2019} who identify substructure through clustering
in integral-of-motion space. 

As shown in \citet{Bird2021}, the two KG samples are very
similar in their properties, and we combine them for our current study.

The final sample sizes are 
10762 K giants and 2526 BHB stars covering a
distance range from $3<r<105$ kpc and $4<r<60$ kpc, respectively. Figures \ref{fig:r}$-$\ref{fig:V} show the distributions for our two smooth, diffuse stellar halo samples in Galactocentric distance, metallicity, and velocity. In this work, to perform the kernel density estimation, we use the {\tt python} package {\tt scipy.gaussian\_kde}; we allow the software to automatically determine the bandwidth which it does so using Scott's Rule \citep{Scott1992}.

We measure the enclosed mass profiles and estimate the virial mass using our KG and BHB samples separately.

  We have sufficiently large samples of both KG and BHB stars to estimate the 3D velocity dispersion and enclosed mass profiles separately.

  We note that our KG and BHB smooth diffuse stellar halo samples peak at different metallicities as seen in the lower panel of Figure \ref{fig:r}. We have shown previously that the metallicities and velocity anisotropy $\beta$ show a relation such that the more metal rich stars have more radial orbits compared to the more metal poor stars \citep{Bird2019beta,Bird2021}. Since our smooth, diffuse KG sample peaks at higher metallicities than our BHB sample, we expect
  the velocity dispersions of the samples to also differ, which we see in Figure \ref{fig:V} where the dispersion is slightly larger in radial velocity for KG stars and in tangential velocities for BHB stars.
  The different velocity distributions will influence our Jeans and TME mass estimations.
Differences in the tracer density profiles will correspond to differences in the velocity dispersion profiles, since the tracer stars occupy a common gravitational potential,  As the uncertainties are large for our estimated density profiles,
we prefer to measure the mass from the KG and BHB samples separately and, as we detail in the following Section \ref{sec:density-prof}, use the density profiles from the literature measured separately for KG and BHB samples. In this way, differences in the Galactic mass profile measured between the KG and BHB tracers will give us an estimation of the mass uncertainties.

\subsection{Data: Tracer Density Profile} \label{sec:density-prof}

\begin{figure}
\includegraphics[width=.95\columnwidth]{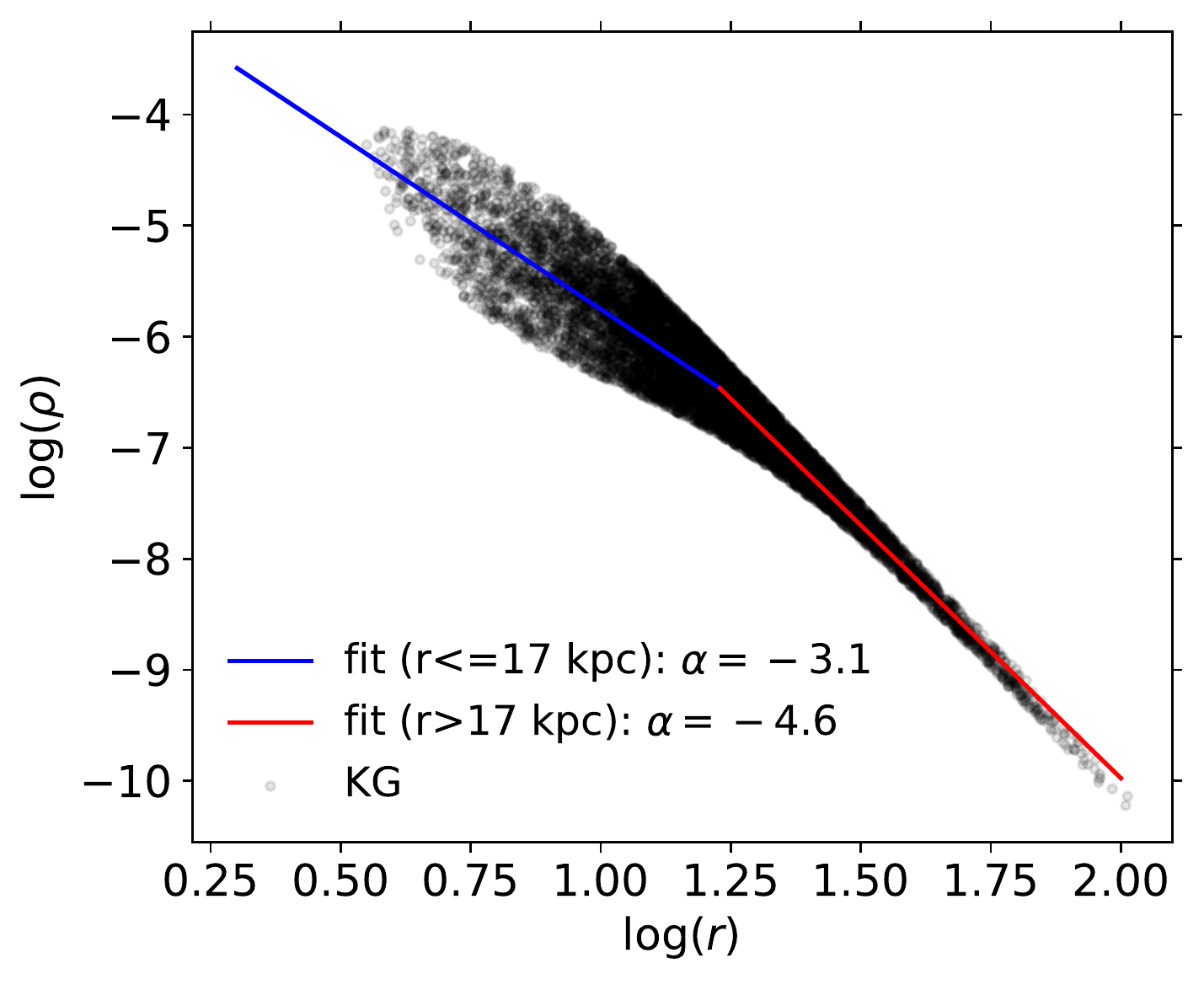}
\includegraphics[width=.95\columnwidth]{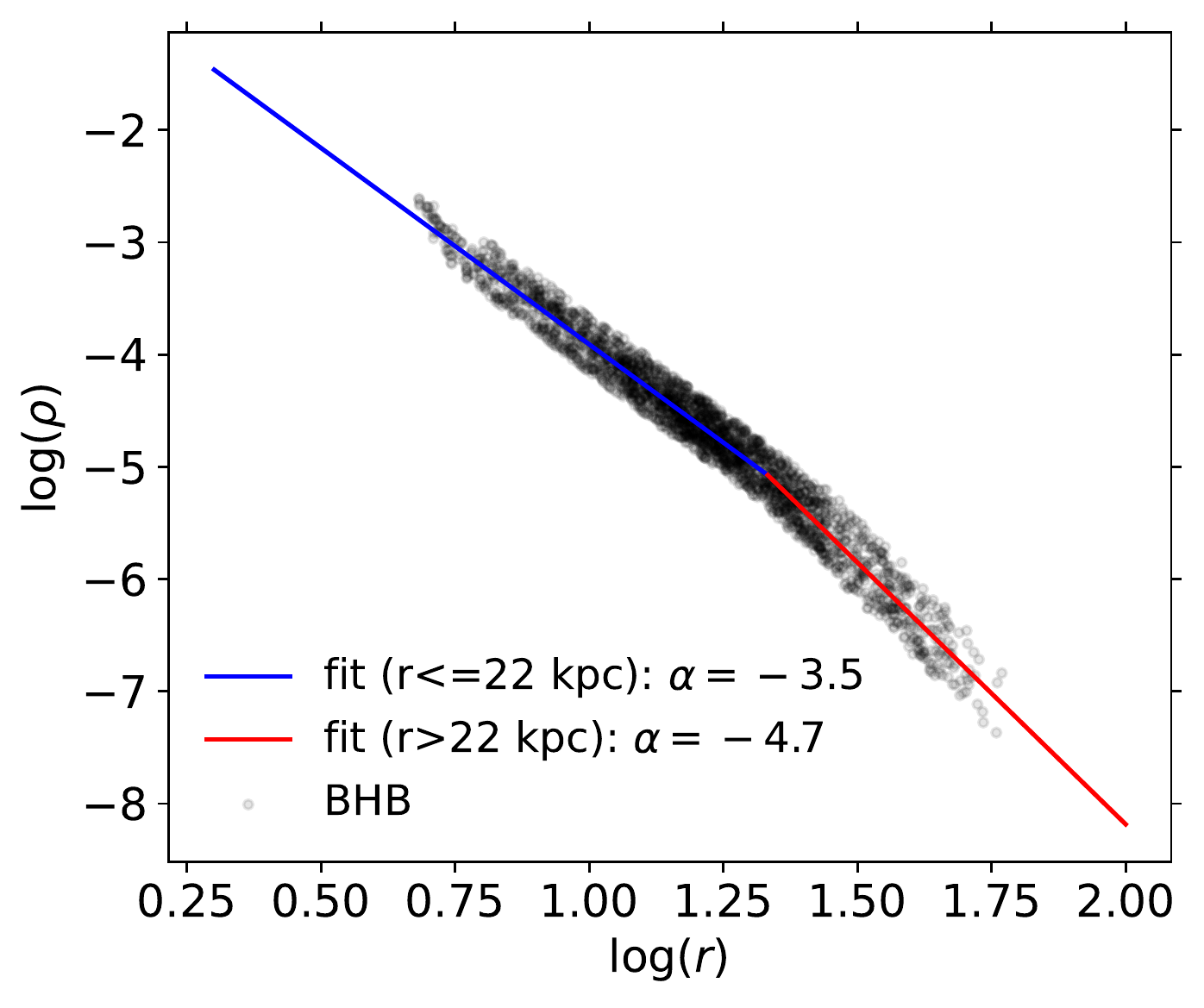}
\caption{Broken power law density $\rho(r)$ profile fits (blue and red lines for inner and outer slopes, respectively) for stars (black symbols) in spherical shells along Galactocentric radius $r$ using the \citet{Xu2018} and \citet{Das2016II} models. Our resulting 
  fits are displayed in the legend. We note that using $\alpha_\mathrm{in}=-3.1$ leads to circular
velocity $>250$ km
s$^{-1}$ near the Sun for our KG sample; in our following mass estimate, we therefore 
adjust our selected value for KG stars to $\alpha_\mathrm{in}=-2.8$
which gives $v_\mathrm{circ}<250$ km s$^{-1}$ near the Sun. 
} 
  \label{fig:density}
\end{figure}

The density fall-off $\rho(r)$ of the
tracer stars is an important assumption in the mass determination,
both for the Jeans equation and TME.

We have chosen to use the stellar density models of
\citet{Xu2018} and \citet{Das2016II} for KG and BHB
stars, respectively, (also see \citet{Xue2015} and \citet{Deason2011.416} for similar density profiles based on KG and BHB stars, respectively).
For the KG stellar halo we use the model which includes flattening of the stellar halo that changes with radius and a single power law density; this is the best model determined by \citet{Xu2018} as compared to other models tested. \citet{Das2016II} use a broken power law density model with constant flattening of the stellar halo.

These authors use similar halo star samples as we use in the current
work, but the density profiles are derived based on a flattened
stellar halo.

Since we use the spherical Jeans equation and TME (which also assumes a spherical system) in this work, we have
computed the density profiles for spherical shells along
Galactocentric radius $r$ using the \citet{Xu2018} and \citet{Das2016II}
models.

We fit a broken power law density model using a Bayesian MCMC maximum likelihood method and make use of the {\tt
    PYTHON} package {\tt emcee} \citep{emcee}.
We use
300 walkers, 700 runs, and 200 runs for the burn-in phase. 
We use the 50th percentile as the best fit and we investigate the goodness-of-fit using the 16th and 84th percentiles as the random uncertainty.

Our resulting broken power law density profile
fits are $\alpha_\mathrm{in}=-3.1$, $r_\mathrm{break}=17$ kpc, and
$\alpha_\mathrm{out}=-4.6$ for our combined LAMOST and SDSS KG stars, and
$\alpha_\mathrm{in}=-3.5$, $r_\mathrm{break}=22$ kpc, and
$\alpha_\mathrm{out}=-4.7$ for SDSS BHB stars.
The uncertainties on $\alpha_\mathrm{in}$ and $\alpha_\mathrm{out}$ are $\sim\pm0.001\sim\pm0.004$ and on $r_\mathrm{break}$ $\pm1$ kpc.
As seen in Figure \ref{fig:r}, over half of our KG and BHB samples are located within their respective $r_\mathrm{break}$.
We show the fitting results in Figure \ref{fig:density}.

Using $\alpha_\mathrm{in}=-3.1$, we find that the circular
velocity near the Sun for our KG sample is $>250$ km
s$^{-1}$. As this somewhat high compared to measurements found in the
literature, we adjust our selected value to $\alpha_\mathrm{in}=-2.8$
which gives $v_\mathrm{circ}<250$ km s$^{-1}$ near the Sun. This effectively sets
a lower limit on $\alpha_\mathrm{in}$. We further discuss uncertainties due to the density profile in Section \ref{subsec:dist-and-den}.

\subsection{Radially Binning the Data} \label{sec:bins}

To calculate the enclosed mass profile by applying the 3D spherical Jeans method and TME, we bin the data in Galactocentric radial bins. Several factors influence the bin selection. We select bin sizes taking into account the distance uncertainty of our KG and BHB samples. The KG sample has a 16 percent relative distance uncertainty and the BHB sample has 10 percent. We thus allow the BHB bins to be up to half as many as the KG bins. We also consider the number density of our samples. Over 60 percent of our KG and BHB samples are located within 20 kpc. When selecting the most distant bins, we check that over 10 stars reside in each bin, although the majority of the most distant bins have over 50 stars. Finally, we select the ends of each bin as $r=15,25,35,45,55,65,$ and 125 for our KG sample and $r=7,8,10,13,20,25,30,35,43,50,$ and 125 kpc for our BHB sample. For plotting purposes, we use the median distance of stars within the bin.

\begin{figure}
\includegraphics[width=.95\columnwidth]{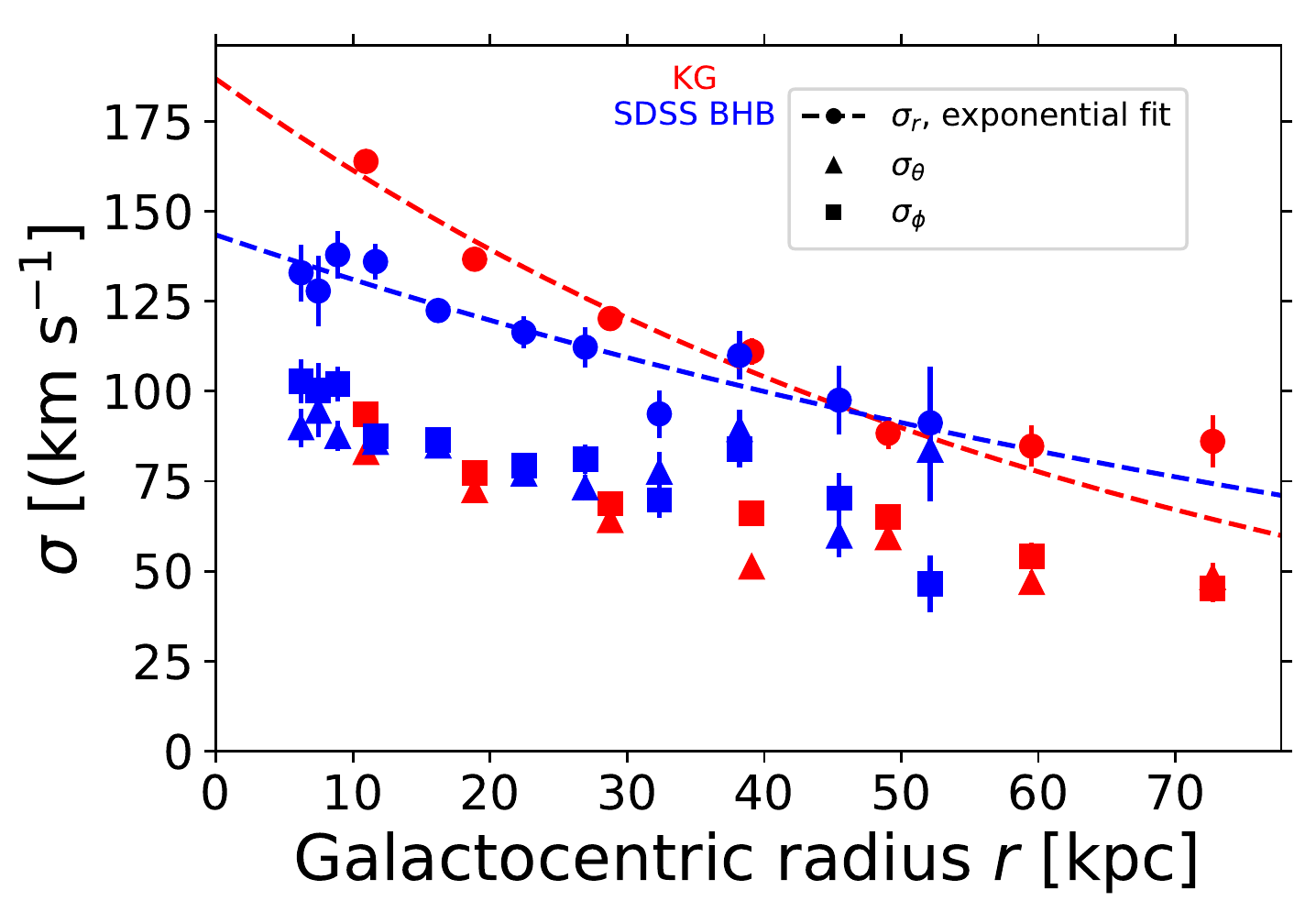}
\includegraphics[width=.95\columnwidth]{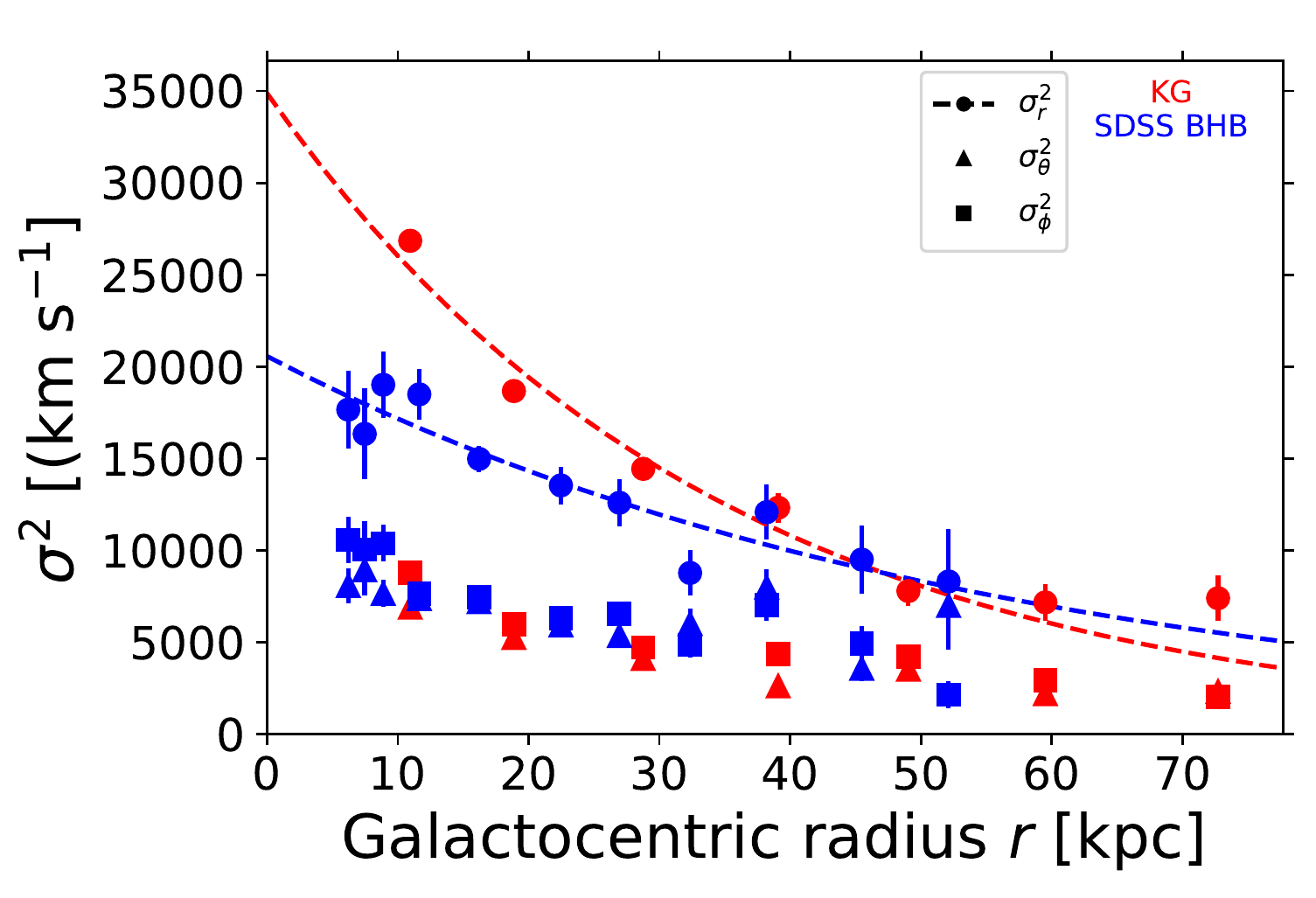}
\includegraphics[width=.95\columnwidth]{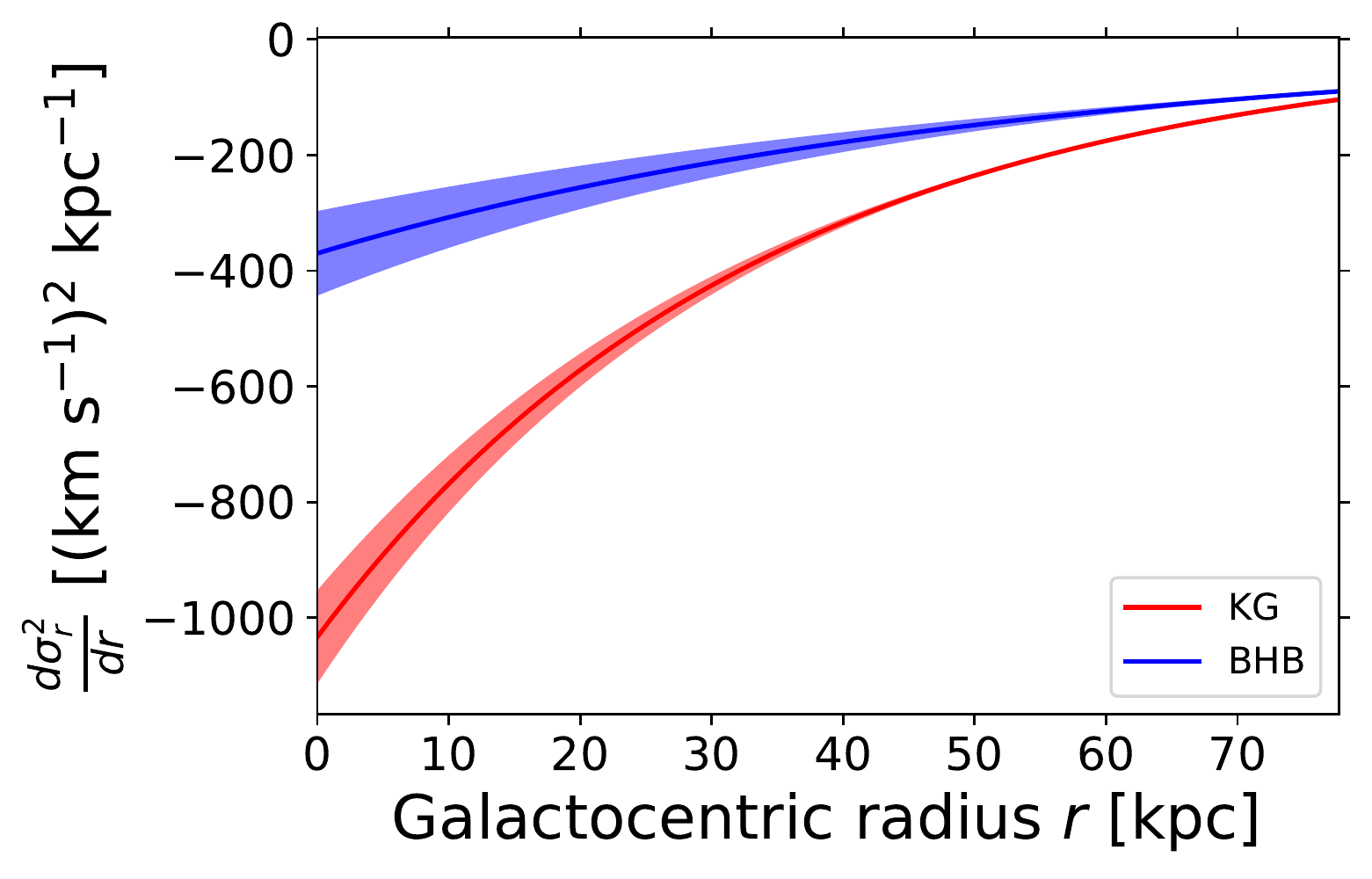}
\caption{Profiles for the 3D spherical $\sigma$ (upper panel), $\sigma^2$ (middle panel), and $\frac{d\sigma_r^2}{dr}$ (lower panel) using KG (red) and BHB (blue) stars. Markers (circle, triangle, square) are $\sigma$ and $\sigma^2$ (in Galactocentric spherical coordinates $r,\theta,\phi$, respectively) measured from the data, whereas dashed lines are the exponential fit for $\sigma_r$.
  The uncertainties propagated from the observed data for each bin (explained in Section \ref{sec:3DJeans}) are included, although many markers are larger than the uncertainty.
  In the lower panel, the shaded regions mark the statistical uncertainty for $\frac{d\sigma_r^2}{dr}$.
  The fit to the $\sigma_r$ velocity term is used to calculate the $\frac{d\sigma_r^2}{dr}$ term that go into our 3D spherical Jeans mass estimate.
Each marker represents the median radius of the stars within our selected radial bins.
} 
  \label{fig:vel-terms}
\end{figure}

\begin{figure}
\includegraphics[width=.95\columnwidth]{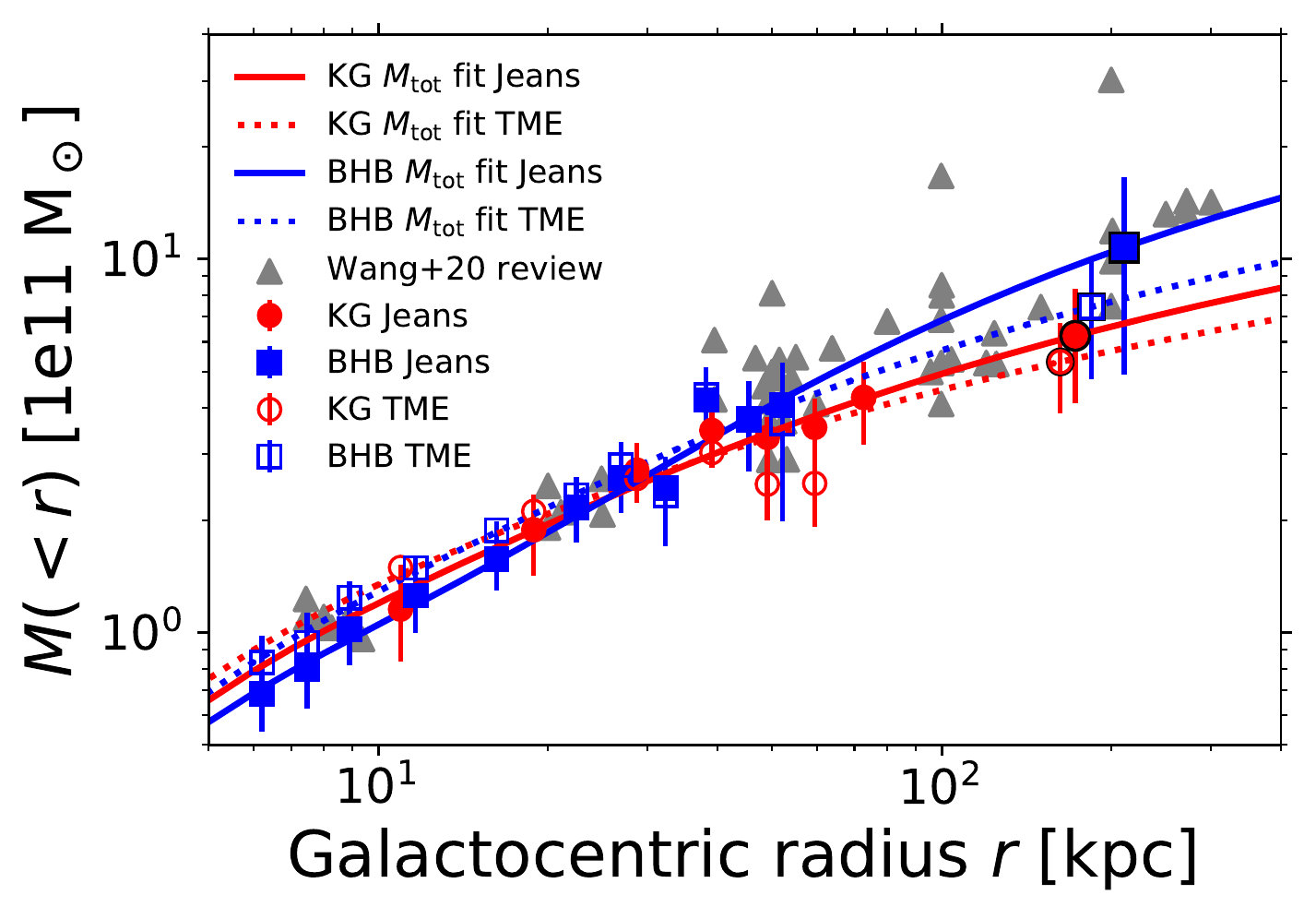}
\includegraphics[width=.95\columnwidth]{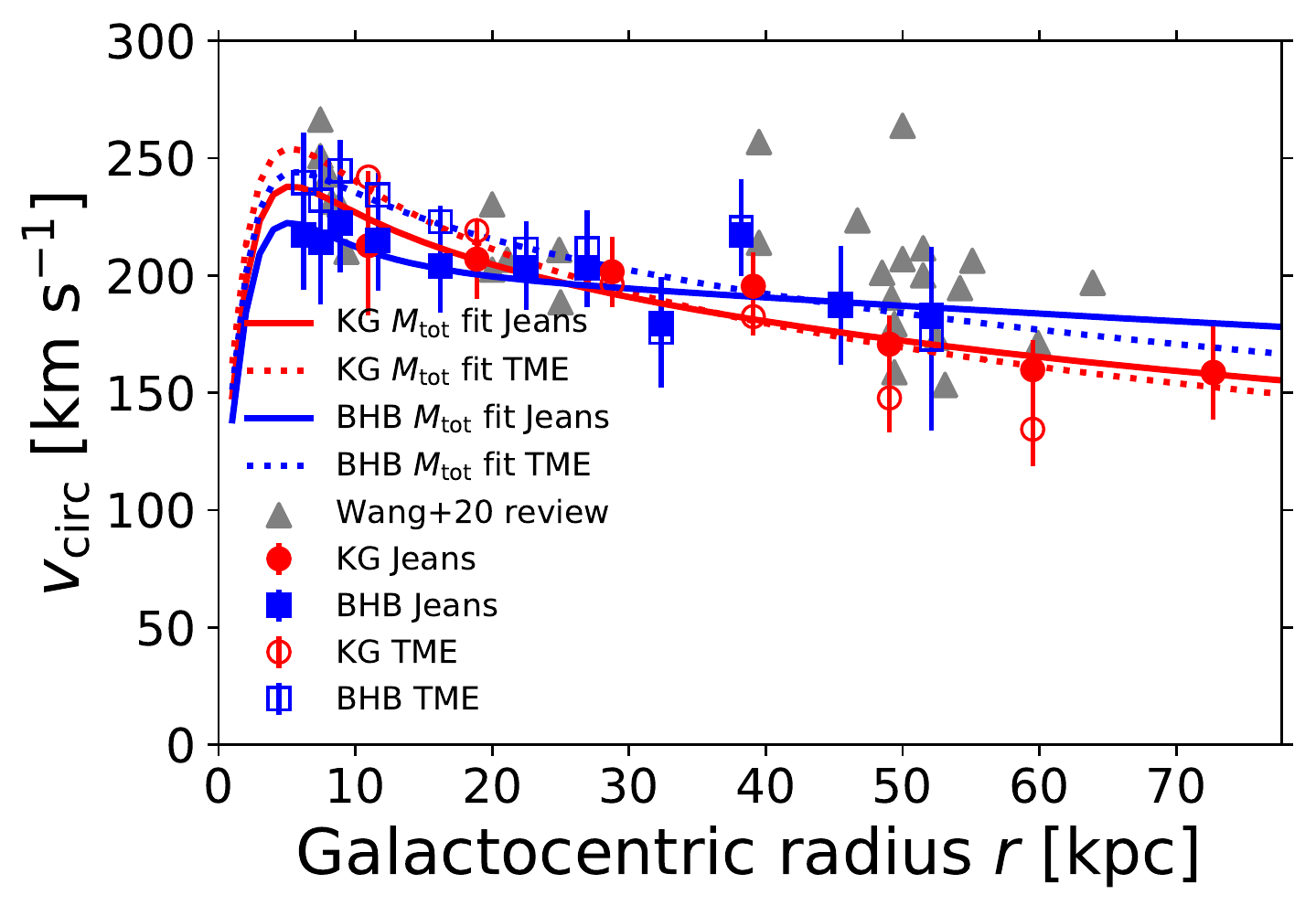}
\caption{Profiles for Milky Way enclosed mass $M(<r)$ (upper panel) and circular velocity $v_\mathrm{circ}$ (lower panel) using KG (red) and BHB (blue) stars.
Jeans (filled markers) and TME (open markers) $M(<r)$ and $v_\mathrm{circ}$ are plot for our KG (circles) and BHB (squares) samples.
Each marker represents the median radius of the stars within our selected radial bins. The $x$-axis of the upper panel is in log scale and of the lower panel linear.
Using the
\citet{Bovy2013,Bovy2015} mass model consisting of a disk and
bulge component and an NFW profile, we fit the data (markers) and plot the models for the Jeans (solid lines) and TME (dotted lines) methods.
The best fit NFW profile yields a dark matter mass $M_{200}$ [$\times10^{12}~\mathrm{M}_\odot$] of
0.55 (KG Jeans), 1.00 (BHB Jeans), 0.46 (KG TME), and 0.67 (BHB TME).
In the upper panel, the most distant markers of each KG (circles) and BHB (squares) mass profile are the corresponding total mass $M_\mathrm{tot}$ estimates within the virial radius $r_{200}$ from the model fits.
Gray triangle markers are the $M(<r)$ and $v_\mathrm{circ}$ estimates from the literature using various methods and are summarized in \citet{Wang_Wenting2020}.
} 
  \label{fig:mass-vcirc-Wang}
\end{figure}

\begin{table}
  \caption{Enclosed total mass estimates.}
\label{table:results}
\begin{center}
\begin{tabular}{ccccc}
\hline
Method (star type) & $<r$\tablenote{Median Galactocentric distance of stars in the last radial bin.} & $M(<r)$\\
 & kpc & $\times10^{11}$ M$_\odot$ \\
\hline
Jeans (KG) & 73 & $4.3\pm0.95$\tablenote{Uncertainties propagated from the observed data are included, as described in Section \ref{sec:3DJeans}. We detail the systematic uncertainty within the text (namely, in Section \ref{subsection:nonvir} we elaborate upon our adopted 15 percent systematic uncertainty due to non-spherical and non-virial effects).} \\
\hline
Jeans (BHB) & 52 & $4.1 \pm 1.2$ \\
\hline
TME (KG) & 73 & $4.3\pm1.1$ \\
\hline
TME (BHB) & 52 & $3.6\pm1.6$ \\
\hline
\end{tabular}
\end{center}
\end{table}

\subsection{Results: 3D Spherical Jeans Milky Way Mass} \label{sec:jeansmass}

\begin{figure}
  \includegraphics[width=.95\columnwidth]{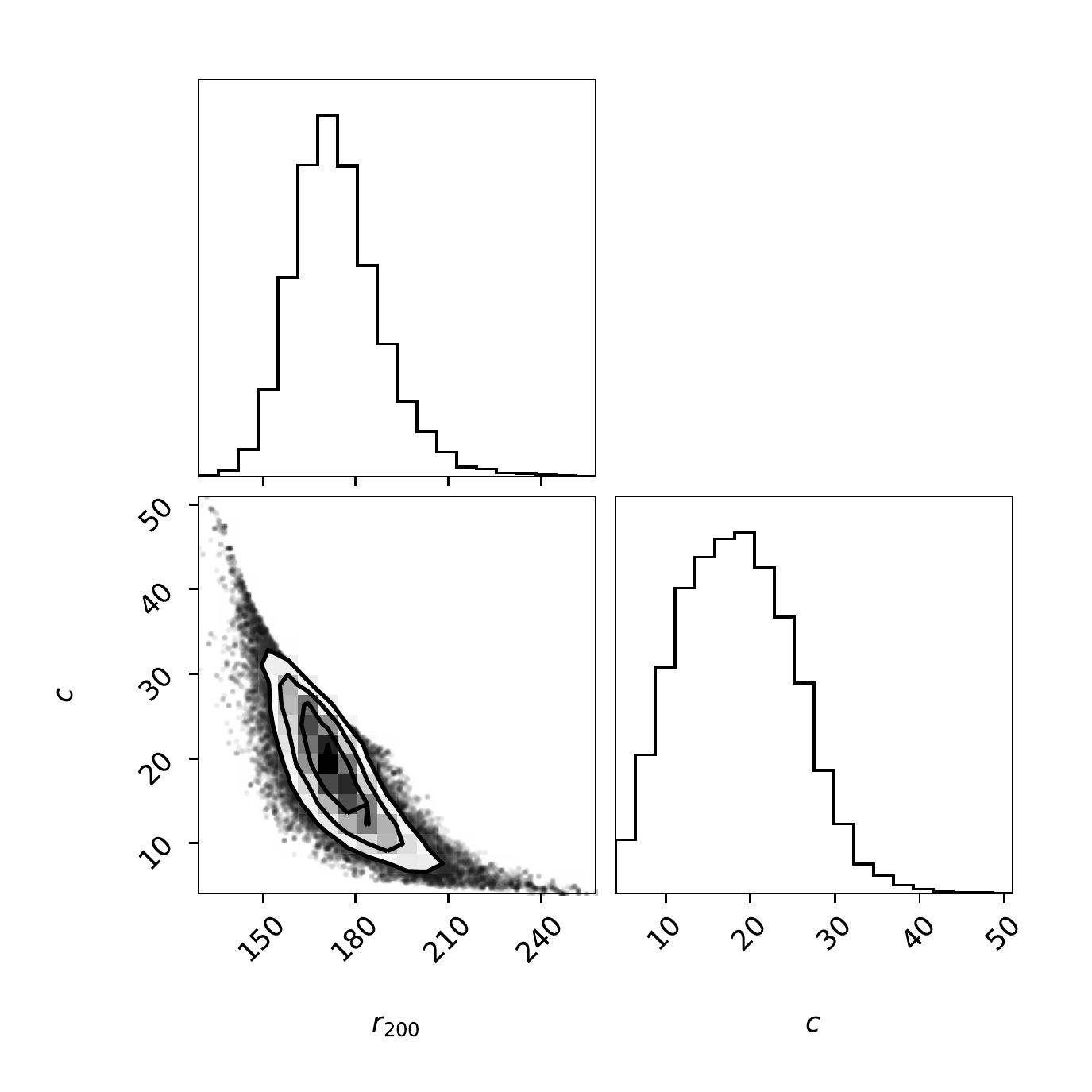}\\
  \includegraphics[width=.95\columnwidth]{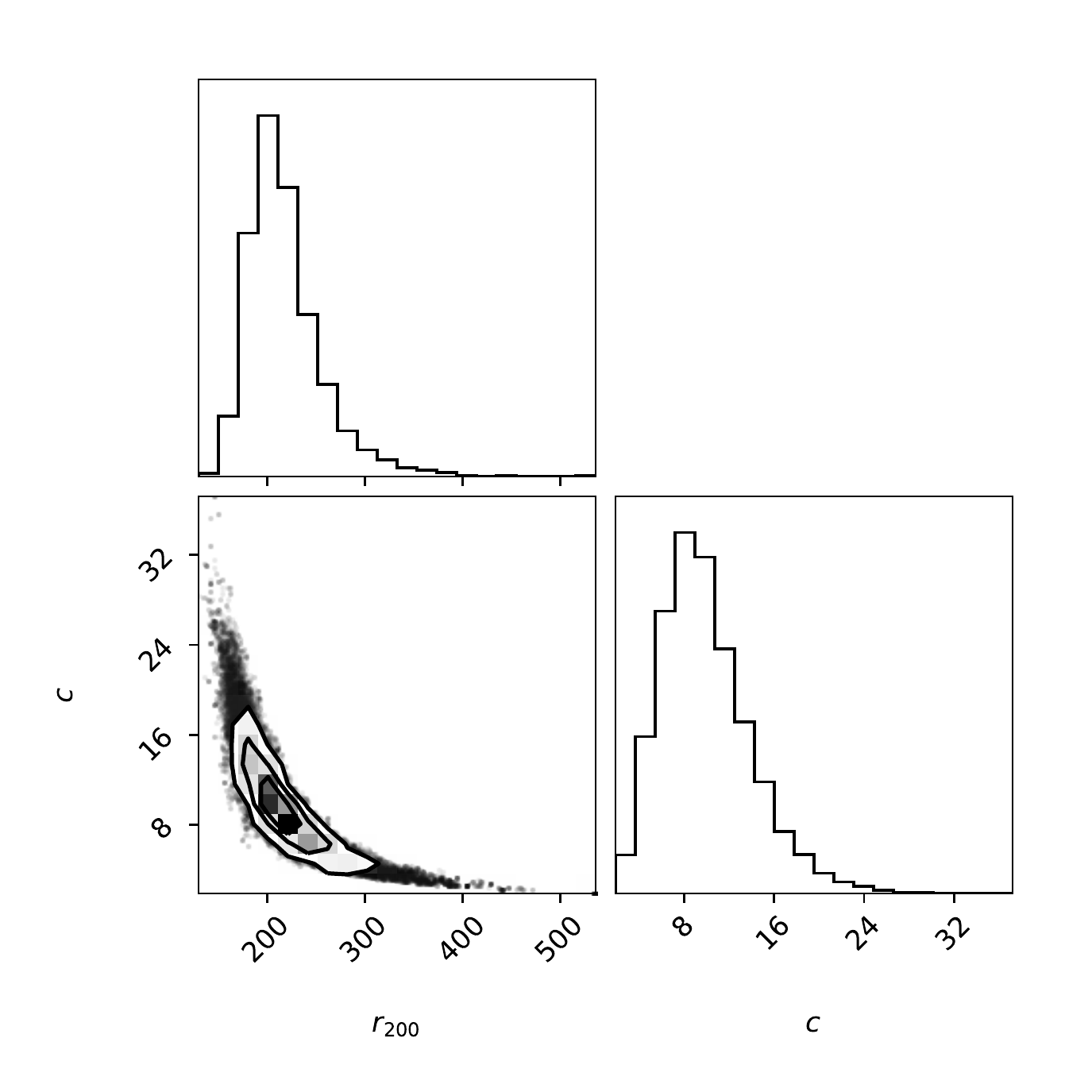}
  \caption{Quartiles for the Bayesian MCMC NFW fitting for $r_{200}$ and $c$ for the KG (upper panel) and BHB (lower panel) 3D spherical Jeans mass profiles in the upper panel of Figure \ref{fig:mass-vcirc-Wang}.
} 
  \label{fig:triangle}
\end{figure}

\begin{figure}
  \includegraphics[width=.95\columnwidth]{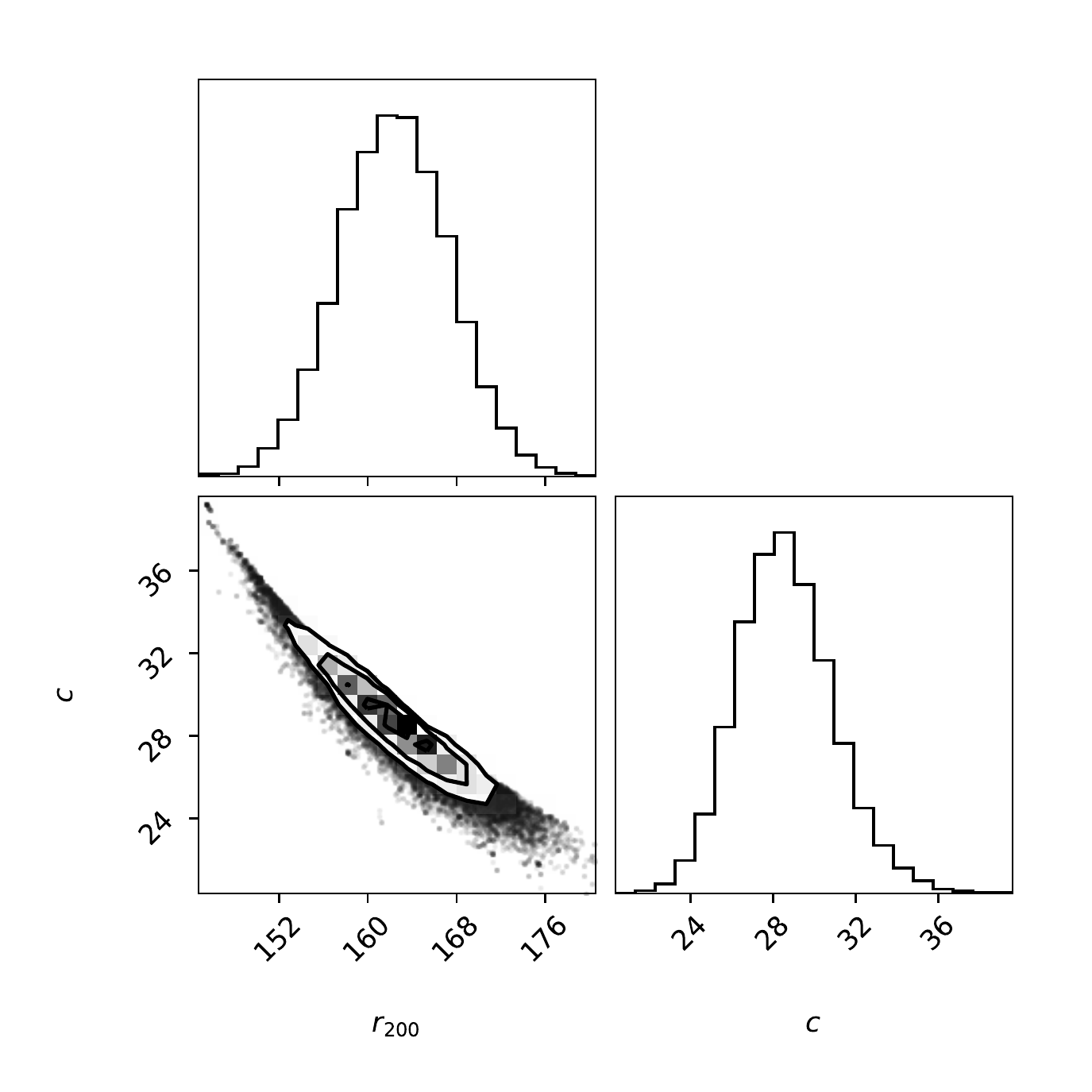}\\
  \includegraphics[width=.95\columnwidth]{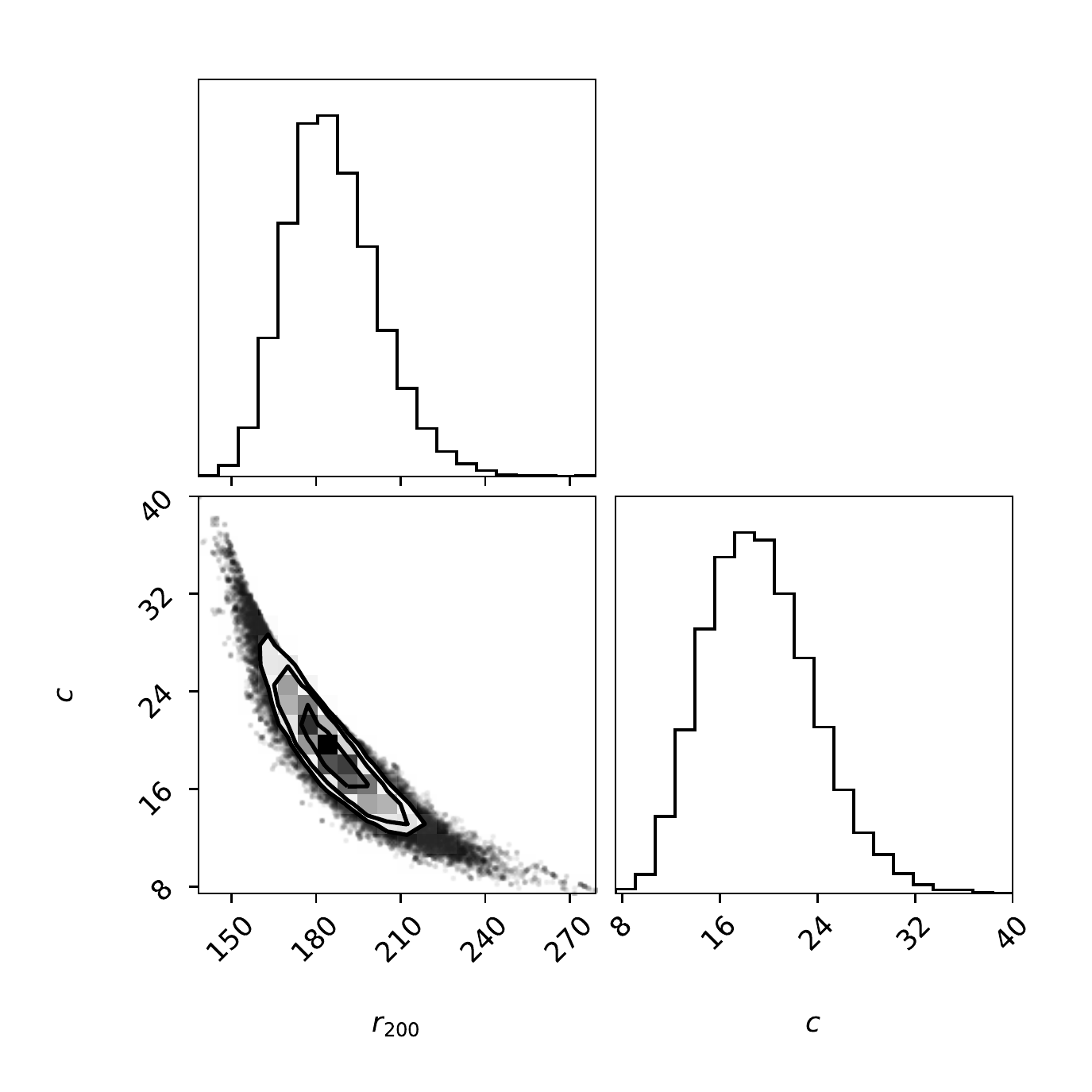}
  \caption{Quartiles for the Bayesian MCMC NFW fitting for $r_{200}$ and $c$ for the KG (upper panel) and BHB (lower panel) TME mass profiles in the upper panel of Figure \ref{fig:mass-vcirc-Wang}.
} 
  \label{fig:triangle-tme}
\end{figure}

To apply the Jeans equation, we first measure the velocity dispersions
$(\sigma_r, \sigma_\phi, \sigma_\theta)$ as functions of
Galactocentric radius $r$ from the sample stars. The velocity
dispersions in the three components $V_r, V_\theta, V_\phi$ have been
computed using the {\tt extreme-deconvolution} algorithm
\citep{Bovy2011}, as described in \citep{Bird2021}. This takes into
account the individual velocity errors on each star in each component,
as well as the covariances between the velocity errors (which are
mostly driven by the covariance between the {\it Gaia} DR2 proper
motions, but also by the distance error for each star).
We use the Poisson statistics of the number of stars in each bin (after using {\tt extreme-deconvolution} to determine the velocity dispersion, taking into account the individual observational errors on the stellar velocities, and the covariances between them).
We detail the propagation of the observational uncertainties in Section \ref{sec:3DJeans}. As mentioned in Section \ref{sec:bins}, we check that over 10 stars reside in each bin, although the majority of the most distant bins have over 50 stars.

In Figure \ref{fig:vel-terms}, we show the velocity terms as
functions of $r$, their uncertainties, and the corresponding exponential fit to
$\sigma_r$. The Bayesian MCMC best fits ($\sigma_0, h_r$, Equations \ref{eqn:s}$-$\ref{eqn:s2}) are $(187\pm3\,\mathrm{km}\,\mathrm{s}^{-1},68\pm3,\mathrm{kpc})$ and $(143\pm5\,\mathrm{km}\,\mathrm{s}^{-1},111\pm20\,\mathrm{kpc})$ for our KG and BHB samples, respectively.

Applying the Jeans
equation yields the enclosed mass $M(<r)$ and circular velocity
($v_\mathrm{circ}(r)=\sqrt{GM(r)/r}$) profiles shown in Figure
\ref{fig:mass-vcirc-Wang}.

Note that the error estimates displayed in Figures \ref{fig:vel-terms}
and \ref{fig:mass-vcirc-Wang} are the
uncertainties propagated from the observational data only (discussed in Sections \ref{sec:3DJeans}$-$\ref{sec:tme}.
These random statistical errors are of order 10 to 30 percent.
As discussed in Section \ref{subsection:nonvir}, we add 15 percent systematic uncertainty to our mass estimate to account for non-virial and non-spherical effects.

We summarize our mass measurements in Table \ref{table:results}.  We
find that the 3D spherical Jeans equation yields the enclosed Milky
Way mass profiles shown in Figure \ref{fig:mass-vcirc-Wang} (upper
panel) with
$M(r<73~\mathrm{kpc})=4.3\pm0.95\times10^{11}~\mathrm{M}_\odot$ and
$M(r<52~\mathrm{kpc})=4.1\pm1.2\times10^{11}~\mathrm{M}_\odot$ for
the KG and BHB samples, respectively.
To these estimates for KG and BHB stars, we add $\pm0.6$ and $\pm0.6$ $\times10^{11}~\mathrm{M}_\odot$, respectively, to account for systematic uncertainties.

We also display in Figure \ref{fig:mass-vcirc-Wang} individual mass estimates from various methods which
are summarized by \citet{Wang_Wenting2020}; specifically, these
include the measurements from \citet{Penarrubia2014},
\citet{Kupper2015}, \citet{Malhan2019}, \citet{McMillan2011,
  McMillan2017}, \citet{Nesti2013}, \citet{Posti2019},
\citet{Watkins2019}, \citet{Kafle2012}, and \citet{Eadie2019}.

\subsection{Results: TME Milky Way Mass} \label{sec:tmemass}

We find that the \citet{Evans2011} mass estimator yields the enclosed
Milky Way mass profiles shown in Figure \ref{fig:mass-vcirc-Wang}
(upper panel) with
$M(r<73~\mathrm{kpc})=4.3\pm1.1\times10^{11}~\mathrm{M}_\odot$ and
$M(r<52~\mathrm{kpc})=3.6\pm1.6\times10^{11}~\mathrm{M}_\odot$ for
the KG and BHB samples, respectively.
We estimate the random uncertainty on the TME mass by bootstrapping the stellar sample as described in Section \ref{sec:tme}.
As with the Jeans method, these random statistical uncertainties are of order 10 to 30 percent.

Taking into account the 15 percent systematic uncertainty (outlined in Section \ref{subsection:nonvir}), we add
$\pm0.6$ ($\pm0.5$) $\times10^{11}~\mathrm{M}_\odot$ to the KG (BHB) mass estimate.

We summarize our enclosed mass measurements in Table \ref{table:results} and plot our mass and circular velocity profiles in Figure \ref{fig:mass-vcirc-Wang}.

\subsection{Results: Virial Mass}
\label{results:m200}

\begin{table}
\caption{Best fit parameters for an NFW dark matter profile.}
\label{table:nfw}
\begin{center}
\begin{tabular}{ccccc}
\hline
Method (star type) & $M_{200}$ & $r_{200}$ & $c$\\
 & $\times 10^{12}$ M$_\odot$ & kpc & \\
\hline
Jeans (KG) & $0.55^{+0.15}_{-0.11}$\footnote{All estimates are from the Bayesian MCMC fitting, we take the 50th percentile as the best fit and the 16th and 84th percentiles as the random uncertainty.} & $173^{+15}_{-12}$ & $18.2^{+7.1}_{-6.8}$\\
\hline
Jeans (BHB) & $1.00^{+0.67}_{-0.33}$ & $211^{+39}_{-26}$ & $9.4^{+4.4}_{-3.2}$\\
\hline
TME (KG) & $0.46^{+0.043}_{-0.040}$ & $162^{+4.9}_{-4.9}$ & $28.5^{+2.4}_{-2.2}$\\
\hline
TME (BHB) & $0.67^{+0.21}_{-0.14}$ & $185^{+17}_{-14}$ & $19.0^{+4.7}_{-4.1}$\\
\hline
\end{tabular}
\end{center}
\end{table}

We can extrapolate our results in the previous sections for the
directly measured enclosed mass (which reaches distances of $\sim50$ to $\sim70$
kpc) out to the virial radius of the dark halo, by subtracting the
baryonic component and fitting an NFW profile to the residual mass
distribution.

Specifically, using the mass model and Bayesian MCMC method described in
Section \ref{sec:method_m200}, we present the best fitting virial
mass, radius, and concentration parameter to the KG and BHB 3D
spherical Jeans and TME mass profiles in Table \ref{table:nfw}. The total mass estimates (baryonic mass plus dark mass $M_{200}$) at the virial radius for our KG and BHB Jeans and TME mass profiles are shown in Figure \ref{fig:mass-vcirc-Wang} (most distant markers for each corresponding profile).
The
quartiles are shown in Figure \ref{fig:triangle} and Figure
\ref{fig:triangle-tme} for the 3D spherical Jeans and TME methods,
respectively. Together our two samples and methods give a weighted average of $M_{200}=0.53 \pm 0.05\times 10^{12}$ M$_\odot$,
$r_{200}=172\pm 5$ kpc, and $c=20\pm3$.

Note that our constraint on $c$ is dominated by our inner data points,
since the inflection point of the NFW profile is the scale radius
$r_\mathrm{s}=r_{200}/c$. This is as expected, as noted by
\cite{Kafle2014}, who show that tracer stars beyond about 40 kpc add
little to our knowledge of the mass and concentration parameter for
the NFW dark halo: rather, the inner data (in the Galactocentric range 5
to 40 kpc) dominate the fitting.

\begin{figure*}
\includegraphics[width=1.8\columnwidth]{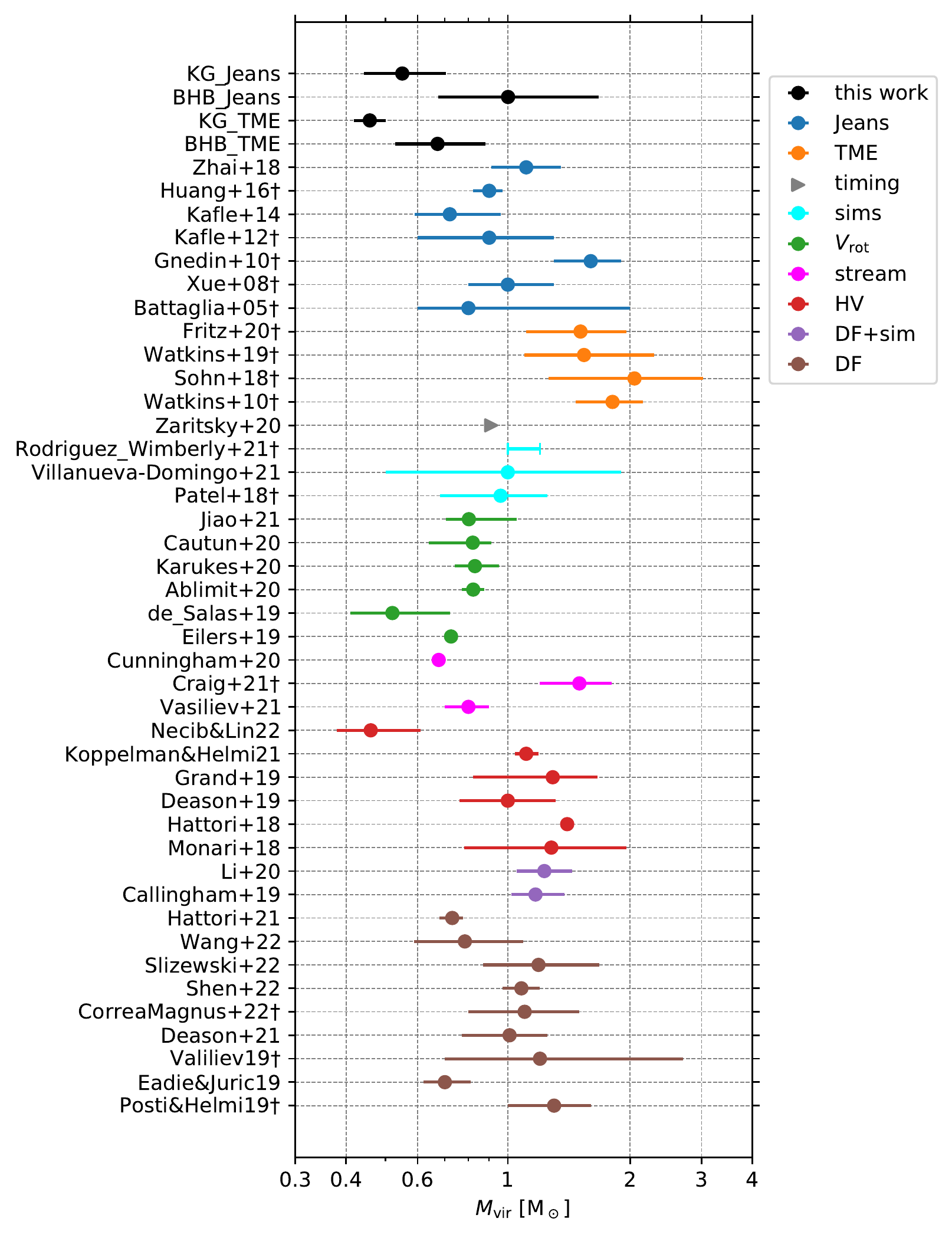}
\caption{Comparison of our KG and BHB dark matter virial mass $M_\mathrm{200}$ estimates with virial mass $M_\mathrm{vir}$ estimates from the literature. We plot those measured using similar Jeans and TME methods involving tracer samples, as well as the recent virial mass estimates using various different methods. Colors represent the method as specified in the legend. The $\dagger$ marks virial mass estimates using definitions other than $M_{200}$. As the following exceptions, we plot the enclosed mass estimate within 200 kpc of \citet{Watkins2010}, the lower limit $M_{200}$ estimate from \citet{Zaritsky2020}, the best estimated range of mass measured by \citet{Rodriguez_Wimberly2022}, and we do not show uncertainties for \citet{Hattori2018.866} and \citet{Cunningham2020} since none are quoted for these mass estimates. The $x$-axis is in log scale. The weighted average virial mass is $\sim0.85\times10^{12}$ M$_\odot$  with a scatter of $<5$ percent; this includes all mass estimates shown excluding \citet{Zaritsky2020}, \citet{Rodriguez_Wimberly2022}, \citet{Hattori2018.866}, and \citet{Cunningham2020} for the reasons above.
}
\label{figmass-compared}
\end{figure*}

We compare our dark matter halo $M_{200}$ values with other virial mass estimates in the literature also
using similar methods involving tracer samples in addition to the most recent mass estimates in Figure
\ref{figmass-compared}. Among the estimates shown which use either Jeans or TME
\citep[namely,][and including our estimates]{Battaglia2005, Xue2008, Gnedin2010, Watkins2010,
  Kafle2012, Kafle2014, Huang2016, Zhai2018, Sohn2018, Watkins2019,
  Fritz2020} the weighted average virial mass is $\sim0.83\times10^{12}$ M$_\odot$ with a
scatter of $<10$ percent. The weighted average virial mass for all those shown in Figure \ref{figmass-compared} is $\sim0.85\times10^{12}$ M$_\odot$  with a scatter of $<5$ percent; this excludes \citet[][lower limit $M_{200}$]{Zaritsky2020}, \citet[][range of best fitting $M_\mathrm{vir}$]{Rodriguez_Wimberly2022}, \citet[][no $M_\mathrm{vir}$ uncertainties]{Hattori2018.866}, and \citet[][no $M_\mathrm{vir}$ uncertainties]{Cunningham2020} for the reasons in parentheses. We include further discussion of Figure \ref{figmass-compared} in Section \ref{sec:compare}.

\section{Discussion and Conclusions} \label{sec:discussion}

\subsection{Differences Between Tracer Types}

As our collection of smooth, diffuse halo KG and BHB stars differ in several sample properties, we prefer to estimate the mass separately using each sample. In this section we elaborate upon the differenes and our reasoning to leave the samples separate.

Differences in the Galactic mass profile measured between the KG and BHB tracers give
us an estimation of the mass uncertainties. We compare the fits of the
total enclosed mass profiles (baryons plus NFW dark matter) for the KG
versus BHB fits.  We find that our mass estimates using the same
method but different tracers typically differ by 15 to 30 percent at a given radius.
This represents an estimate our systematic errors since the samples
for the different tracers differ in a number of ways.

As mentioned in Section \ref{sec:data}, we compare the distance, metallicity, and velocity distributions of our KG and BHB samples in Figures \ref{fig:r}$-$\ref{fig:V}. Although our KG sample has $\sim4$ times more stars than our BHB sample, we see from the cumulative probability and one dimensional kernel density estimation of Figure \ref{fig:r} that the distribution of distance is quite similar for both samples. The cumulative probability shows that more than half of the samples are located within the break radius of the power law density which we show in Figure \ref{fig:density}. On the other hand, the two samples differ in their metallicity and velocity distributions as seen in the lower panel of Figure \ref{fig:r} and in Figure \ref{fig:V}. The KG sample is characterized with higher metallicity and radial velocity dispersion compared to the BHB sample which peaks at lower metallicities and slightly less radial velocity dispersion.

  Correlations between the velocity and metallicity distributions have recently been noted for various stellar halo samples \citep[e.g.,][]{Myeong2018action,Deason2018,Belokurov2018,Lancaster2019,Bird2019beta,Bird2021}.
  This correlation is largely a consequence of the ancient $Gaia$-Enceladus-Sausage merger. As shown by \citet{Wu2022}, as high as $\sim30-60$ percent of our smooth, diffuse halo samples were deposited by the ancient satellite; the member orbits have long since virialized in our Galaxy's potential and are not readily selected as substructure in integrals of motion by the \citet[][in preparation]{Xue2019} method we use.

  We find that the outer density power law slopes for our KG and and BHB samples are similar ($\alpha_\mathrm{out}=-4.6$ and $-4.7$, respectively). The break radius is slightly further out for the BHB sample ($r_\mathrm{break}=17$ and 22 kpc for our KG and BHB samples, respectively). We find that in order to maintain our criteria of $v_\mathrm{circ}<250$ km s$^{-1}$, the KG sample requires a more shallow inner density power law slope than we recover from our fit to the data and compared to the BHB sample; we must increase $\alpha_\mathrm{in}$ from $-3.1$ to $-2.8$, both of which are more shallow than that of the BHB sample $\alpha_\mathrm{in}=-3.5$.

Studies also find that the flattening of the density distribution of the stellar halo (deviation from sphericity) shows a dependency with metallicity \citep[e.g.,][]{Chiba2000,Carollo2007,Sato2022} such that the more metal rich stellar halo samples are more flattened compared to the more metal poor. As our KG sample peaks at higher metallicities, they likely represent a more flattened stellar halo system than our BHB sample. This is important because in our application of the Jeans equation and TME, we assume that the tracer sample is a spherical system tracing a spherical dark matter halo. If our KG sample if more flattened (an investigation which is beyond the efforts of this current study), our mass estimates from the KG sample will suffer from a slightly different bias compared to estimates using the BHB sample. As we elaborate upon in Section \ref{subsection:nonvir}, we add a 15 percent systematic uncertainty which accounts for, e.g., non-sphericity of the density distribution of our halo tracer sample.

The LAMOST K
giants are selected on the basis of effective temperature and surface
gravity estimates \citep{Liu2014} which are based on LAMOST spectra. 
The SDSS/SEGUE K giants are a collection of three main KG samples with slightly different spectroscopic selection criteria \citep[see, e.g.,][]{Xue2014}.
The BHB stars are selected
similarly, but via very different spectral features (predominantly due
to hydrogen, whereas the K giants predominantly use Mg lines, and
molecular bands due to TiO, CN, and CH).

BHB stars have smaller distance uncertainties than the KG stars; their absolute
luminosities are much easier to calibrate, because of the nearly
constant luminosity of the horizontal branch, compared to the wide
range of luminosities covered by K giants.
On the other hand, the number of BHB stars in our sample is fewer than K giants, so the sampling uncertainties are larger for our BHB sample. 

The BHB stars are only the blue part of the horizontal branch; the red
horizontal branch is better represented by redder stars such as RR
Lyrae. Although the K giants have larger distance uncertainties compared to BHB
stars, they represent a more complete sample of old
populations because they later evolve to both blue and red 
parts of the horizontal branch.

BHB stars are selected spectroscopically, strongly relying on Balmer line cuts, thus they may suffer
from larger selection effects than K giants as the large majority of our KG sample are selected by effective temperature and surface gravity criteria. As shown by
\citet{Liu2014}, their proposed KG selection criteria (the method
we use for our LAMOST sample) select red-giant-branch stars over their full
range of metallicities, and the contamination by stars other than K
giants is small, $<2.5$ percent.

Taking these into consideration, we expect that the KG sample is
more complete and the better sample to measure the stellar halo
density, kinematics, and mass profiles in future efforts.

\subsection{Comparison of 3D Spherical Jeans and TME}

The 3D spherical Jeans equation requires an assumption about the form
of the potential, i.e., that it is spherical, that the velocity
distribution can be well modeled with a single 3D velocity ellipsoid,
and that the effects of bulk velocity flow are negligible.

The TME
makes these assumption as well, and two additional assumptions for simplicity, that the
dark matter halo
follows an NFW profile and that the gravitational potential is well-described by a power law $\gamma$.
\citet{Watkins2010} find the gravitational potential for NFW halos are well-described by a power-law of $\gamma=0.5$, which is what is assumed in the \citet{Evans2011} TME that we use in this work.

Any difference between the Jeans method and the TME will likely be
ascribed to these two assumptions.

Generally speaking in any analysis, the fewer the assumptions the
better, and in this sense the Jeans equation is the preferred of our
two methods. Additionally, it has significant advantages over the TME
as we make full use of the 3D velocity distribution of the tracers (in
particular the transverse velocities), whereas the TME parameterizes
the behavior of the transverse velocities into the scaling factor
using $\beta$.

On the other hand, in our application of the Jeans equation we assume a functional form to the radial velocity dispersion,
which is entirely obviated in
the TME method.
The gradient of the radial velocity dispersion profile
is required by the Jeans equation, and, although direct differentiation can be used, we have chosen a smoothly
differentiable fit to this profile in order to compute this term.
We choose an exponential function since it is easily differentiable and well behaved at the limits of the data.
We explored linear and spline fits as well, and in several recent works, for example, B-splines have been used \citep{Rehemtulla2021,Rehemtulla2022}.

\subsection{Systematic Uncertainty in Distance and Density of Tracers}
\label{subsec:dist-and-den}

Systematic uncertainty in the distances of the tracer stars leads to
systematic error in the enclosed mass estimate: this is
straightforward to estimate. Less straightforward is the form of the
tracer density profile, which we here model as a power-law fall-off.

We show in \citet{Bird2019beta}, that the systematic errors on the
distances to the K giants and BHB stars is of order 10 percent. This
affects the density profile fall-off, but also the transverse
velocities of the stars through the proper motions (particularly at
large Galactocentric distances). In \citet{Bird2019beta} and
\citet{Bird2021} we show the effects on velocity anisotropy due to 10
percent distance uncertainties in our samples. We find that a $\pm 10$
percent systematic distance uncertainty causes systematic
uncertainties in the mass estimates of 20 percent. For full details,
see the Appendix.

More problematic is what uncertainty to adopt for the power-law
fall-off index $\alpha$ in the density of the tracer stars. We are
guided by the significant range of $\alpha$ in the literature, with
values populating the range $-2 < \alpha < -5$, depending on the
methods used and the Galactocentric radii probed.

Additionally, there is good evidence for a break in the power-law
slope at around 20 kpc from the Galactic center. These we denote
$\alpha_\mathrm{in}$ and $\alpha_\mathrm{out}$ within and without this
break radius respectively.

We constrain the uncertainties on $\alpha_\mathrm{in}$ by requiring
that the enclosed mass at 10 kpc must lie in the range of 0.9 to 1.5
$\times\,10^{11}\mathrm{M}_\odot$ in order to be consistent with a
circular velocity of $200<v_\mathrm{circ}(10\,\mathrm{kpc})<250$ km s$^{-1}$ (we choose a larger
uncertainty range compared to that measured by \citet{Bovy2012} to
allow for a generous budget of possible values for $\alpha$). This
leads to an uncertainty of 20 percent error in the mass due to
uncertainty in $\alpha_\mathrm{in}$.  As mentioned earlier,
observational constraints on the slope of the outer power-law fall-off,
$\alpha_\mathrm{out}$ are quite poor and could easily be as large as
$\pm 1$. Adopting this uncertainty for $\alpha_\mathrm{out} = -4.6 \pm
1$ yields an uncertainty of 20 percent on the enclosed mass at the limit of
the data ($\sim 70$ kpc). This is similar to the uncertainty in the
enclosed mass of 27 percent at 100 kpc found by \citet{Deason2021} (see
their Equation 6) for the same uncertainty of $\pm 1$ in the power-law
slope of the outer halo.

The observational errors in $v_\mathrm{los}$ and proper motions add
negligible uncertainty to the mass estimate compared to other sources
of error.

\subsection{Comparison with Previous Studies}
\label{sec:compare}

The tracer mass estimator of \citet{Evans2003} has been tested by
several authors using simulations
\citep[e.g.][]{Wilkinson1999,Yencho2006,Watkins2010,Deason2011.415}.
\citet{Yencho2006} find a fundamental limit on the accuracy of the
tracer mass estimator of 20 percent, due to systematic errors from
substructure in the tracer population, regardless of sample size and
the accuracy of the parameter measurements.

\citet{Wang2017} and \citet{Han2016b} have analyzed the stellar halo
component of Milky Way type galaxies found in cosmological
simulations. They use the \citet{Han2016a} steady-state spherical
system and find an intrinsic systematic bias in the mass measurement
of up to 30 percent, which is caused by the lack of equilibrium of the
Milky Way size halos.  This large error exists before observational
error or uncertainty in the state of the system or potential are
included.

\citet{Wang2015,Wang2016erratum} find that as Galactocentric distance
increases, mass estimators become increasingly less accurate and the
leading source of error differs from halo to halo.  They conclude the
mass is best constrained within $0.2r_{200}$ from the studied
simulations.  For the Milky Way, the mass estimate even at this radius
may be complicated as we know the Sagittarius Stream is strongly
prominent in the halo star counts in the range $30<r<50$ kpc.

\citet{Wang2015,Wang2016erratum} investigate many of the possible
biases involved with the tracer mass estimator such as the one used in
the current study: correlation between model parameters, deviation
from the NFW model, and violation of model assumptions including
dynamical equilibrium, spherical symmetry,
and infinite halo boundaries.

\citet{Rehemtulla2022} have developed a novel application of the 3D spherical Jeans method, for
the specific case of the Galactic stellar tracer density and velocity distributions modeled
non-parametrically using B-splines. They test their method on stellar halos in Milky
Way-type galaxies in cosmological simulations, incorporating typical observational
uncertainties for current surveys, and are able to recover the enclosed mass profiles with
accuracies of around 20 percent or better depending on the amount of substructure or streaming motion in the galaxy. This uncertainty is largely systematic.

In contrast to their method, our approach is to partly parameterize the fitting.
We do not measure the density profile of the tracer stars from our sample directly,
due to the insufficient sample size and the many selection bias corrections that would be needed to do so.
Instead, we have modeled our KG and BHB stellar density profiles using broken power laws based on
the extensive work of \citet{Xu2018} and \citet{Das2016II} for KG and BHB stars in the halo.
As the Jeans equation requires the derivative of the radial velocity dispersion profile,
we chose to fit it parametrically (using an exponential) to $\sigma_r$ since the function is easily differentiable
and provides an adequate fit of the data. Finally, we use the binned velocity dispersion profiles directly
in the Jeans Equation -- no parametric fitting is required. We thus have a hybrid approach to the enclosed mass determinations.
We are able to achieve mass estimates with random uncertainties of order 20 percent; our uncertainties are not systematic as in those measured by \citet{Rehemtulla2022}.

Our results for the halo mass and concentration compare favorably with
\cite{Kafle2014}, who found $M_\mathrm{vir} = 0.80^{+0.31}_{-0.16} \times
10^{12} $ M$_\odot$ and $c = 21.1^{+14.8}_{-8.3}$, and another similar
recent study by \cite{Huang2016}, who find $M_\mathrm{vir} = 0.9 \pm 0.1$
M$_\odot$ and $c = 18 \pm 1$. We find similar uncertainties to those
found by \cite{Kafle2014}, as we have followed a similarly careful
analysis of random and (particularly) systematic errors.

Our mass estimate is in good accord with the mass estimates of
previous studies applying similar 3D spherical Jeans methods to halo stars \citep[e.g.][note that these
  authors define differently the virial mass]{Battaglia2005, Xue2008,
  Gnedin2010, Kafle2012, Kafle2014, Huang2016}. These mass estimates
are compared in Figure \ref{figmass-compared}.

\citet{Zhai2018} have measured the mass profile to 120 kpc using some
9000 halo K giants from LAMOST, in a study similar to our own. They
use the Jeans equation to derive mass profiles for the Milky Way under
assumptions about the velocity isotropy/anisotropy of the tracer stars. For NFW
dark matter profiles, they derive a virial mass of
$M_{200}=1.08^{+0.17}_{-0.14} \times 10^{12}$ M$_\odot$, $r_{200}=205^{+10}_{-9}$ kpc, and a concentration
parameter of $c = 18.5^{+3.6}_{-2.9}$, which is comparable within the uncertainties with our
estimates for our 3D spherical Jeans KG mass profile with $M_{200}=0.55^{+0.15}_{-0.11} \times 10^{12}$ M$_\odot$ and
best fit values of an NFW profile of $r_{200} = 173^{+15}_{-12}$ kpc and $c =
18.2^{+7.1}_{-6.8}$, as shown in Figure \ref{figmass-compared}.

Most recently \citet{Sohn2018}, \citet{Watkins2019}, and \citet{Fritz2020} have applied the 3D TME method of \citet{Watkins2010} to various data sets with high quality proper motions. \citet{Sohn2018} use globular clusters with {\it Hubble Space Telescope} proper motions. \citet{Watkins2019} and \citet{Fritz2020} use globular clusters and satellites, respectively, with {\it Gaia} proper motions. Our mass estimates are in agreement with these three recent studies also using tracers (of different types) with 3D velocities and a TME method. We include these works for comparison in Figure \ref{figmass-compared}.

  With the results from $Gaia$ and the large number of stars collected from dedicated surveys, a plethora of Galactic mass estimates have recently been made using using different methods and data samples. We summarize these most recent results in Figure \ref{figmass-compared}. 
  Grouping together the works applying a distribution function method 
  \citep{Callingham2019,Posti2019,Eadie2019,Vasiliev2019,Li_Zhao-Zhou2020,Deason2021,Hattori2021,Correa_Magnus2022,Shen2022,Slizewski2022,Wang_Jianling2022}
the weighted mean mass estimate is $\sim0.93 \pm 0.047\times 10^{12}$ M$_\odot$.
The methods using high velocity stars with full 6D phase-space information have been used to estimate the mass of the Milky Way \citep{Monari2018,Deason2019.485,Grand2019,Koppelman2021,Necib2022b} have a weighted mean mass of $\sim0.93 \pm 0.099\times 10^{12}$ M$_\odot$.
Two recent works using either the Sagittarius or Magellanic Steams 
\citep{Vasiliev2021,Craig2022} give a weighted mean mass of $\sim0.98 \pm 0.15\times 10^{12}$ M$_\odot$.
The lightest mass measurements have come from rotation curve based methods \citep{Eilers2019,de_Salas2019,Ablimit2020,Karukes2020,Cautun2020,Jiao2021} with a weighted mean of $\sim0.76 \pm 0.023\times 10^{12}$ M$_\odot$.
The works which use 6D satellite phenomenology compared to Milky Way-type simulations \citep{Patel2018,Villanueva-Domingo2022} infer a weighted mean Galactic mass of $\sim0.97\pm 0.0091\times 10^{12}$ M$_\odot$
and \citet{Rodriguez_Wimberly2022} find their results are consistent with a Galactic mass in the range of $1-1.2\times10^{12}$ M$_\odot$. \citet{Zaritsky2020} apply the timing argument to distant Milky Way halo stars to derive a lower limit to the Milky Way mass of $0.91\times10^{12}$ M$_\odot$. Within the uncertainty, our BHB mass estimates are in agreement with \citet{Zaritsky2020}, although our KG mass estimates and a few of the recent mass estimates which we plot in Figure \ref{figmass-compared} are lower than this limit.

In summary, we use two mass estimate methods on LAMOST and SEGUE K giants, 
and on SDSS BHB stars, finding good agreement between the methods and samples within the limits
of our methods such as the systematic uncertainty in the tracer density profile.

The rotation curve we find is also in good agreement with
the \citet{Huang2016} rotation curve measured from a mixed sample of
H$~$\small I \normalsize gas, disk primary red clump stars and SEGUE
halo K giants. We find our rotation curve is likewise in good agreement
with the recent works of \citet{Eilers2019}, \citet{Mroz2019}, and \citet{Ablimit2020} who use various tracer stars in the disk, including red giants and Cepheids.

A dark matter virial mass of $M_{200}=0.53 \pm 0.05\times 10^{12}$ M$_\odot$ 
(where we have taken the weighted average of the values in Table
\ref{table:nfw}), when compared to the Milky Way's total baryonic mass of
$8\times10^{10}$ M$_\odot$ \citep[][in this review of Milky Way properties, they include the stellar disk, bulge, cold disk gas, and hot halo gas in their total baryonic mass budget]{Bland-Hawthorn2016},
yields a total baryonic mass fraction for the Milky Way of 0.15.

Systematic errors still dominate the uncertainty
budget. In this study, the dominant one is the uncertainty in
the density profile of the tracer stars.
The
immediate future looks bright however, as
improved analyses of the stellar halo density profile are 
being made (see, e.g., \citet{Thomas2018} and references therein).
Improved mass estimates for the Milky Way
halo will result from these efforts in the near future.
In two very interesting studies, \citet{Kafle2018jeans} and \citet{Wang2018}
have shown that with appropriate tracers, the accuracy and
precision of mass profile determinations for Milky Way-type
galaxies (in simulations) can be constrained to better than 12 percent
via the Jeans equation.

\subsection{Conclusions} \label{sec:conclusions}

We use the 3D spherical Jeans equation and the \citet{Evans2011} TME
to measure the Milky Way mass profile out to a Galactocentric distance
of $\sim50$ and $\sim70$ kpc with the smooth, diffuse stellar halo samples of LAMOST and 
SDSS/SEGUE K giants and SDSS/SEGUE BHB
stars from \citet{Bird2021}. The extent of this
sample reaches to past 100 kpc.  For our two star types and their
corresponding mass profiles derived from the two methods, we find a weighted
average dark matter virial mass of
$M_{200}=0.53 \pm 0.05\times 10^{12}$ M$_\odot$ with weighted
average virial radius and concentration parameter
$r_{200}=172\pm 5$ kpc and $c=20\pm3$.

\acknowledgments

We thank the anonymous referee for comments and suggestions which have improved the clarity and thoroughness of the manuscript. This work is supported by the National Key R\&D Program of China No. 2019YFA0405500; National Natural Science Foundation of China under grants No. 11988101, 11890694, 11873057, 11873052, and 11873034; China Manned Space Project with No. CMS-CSST-2021-B03, CMS-CSST-2021-A09, and CMS-CSST-2021-A08; and Hubei Provincial Outstanding Youth Fund No. 2019CFA087. S.A.B. acknowledges support from the Aliyun Fellowship, Chinese 
Academy of Sciences President's International Fellowship
Initiative Grant (No. 2016PE010 and 2021PM0055), and
Postdoctoral Scholar's Fellowship of LAMOST.

\software{{\tt Astropy} \citep[v4.2.1;][]{astropy2013,astropy2018}, {\tt emcee} \citep[v3.1.1;][]{emcee}, {\tt extreme-deconvolution} \citep{Bovy2011},
  {\tt galpy} \citep[v1.6.0.post0;][]{Bovy2015}, {\tt matplotlib.pyplot} \citep[v3.3.4;][]{Hunter2007}, {\tt NumPy} \citep[v1.20.1;][]{Oliphant2006,Walt2011,Oliphant2015,Harris2020array}, {\tt pyia} \citep{Price-Whelan2018}, {\tt SciPy} \citep[v1.6.2][]{Virtanen2020}, {\tt TOPCAT} \citep[v4.8-2;][]{Taylor2005}.
}

{\it Data Availability Statement:}\\
The data presented in the figures are available upon request from the
authors.

\bibliographystyle{apj} 
\bibliography{Bird_2022_sjmass}


\appendix
\label{appendix}

\begin{figure}
\begin{tabular}{cc}
\includegraphics[width=.5\columnwidth]{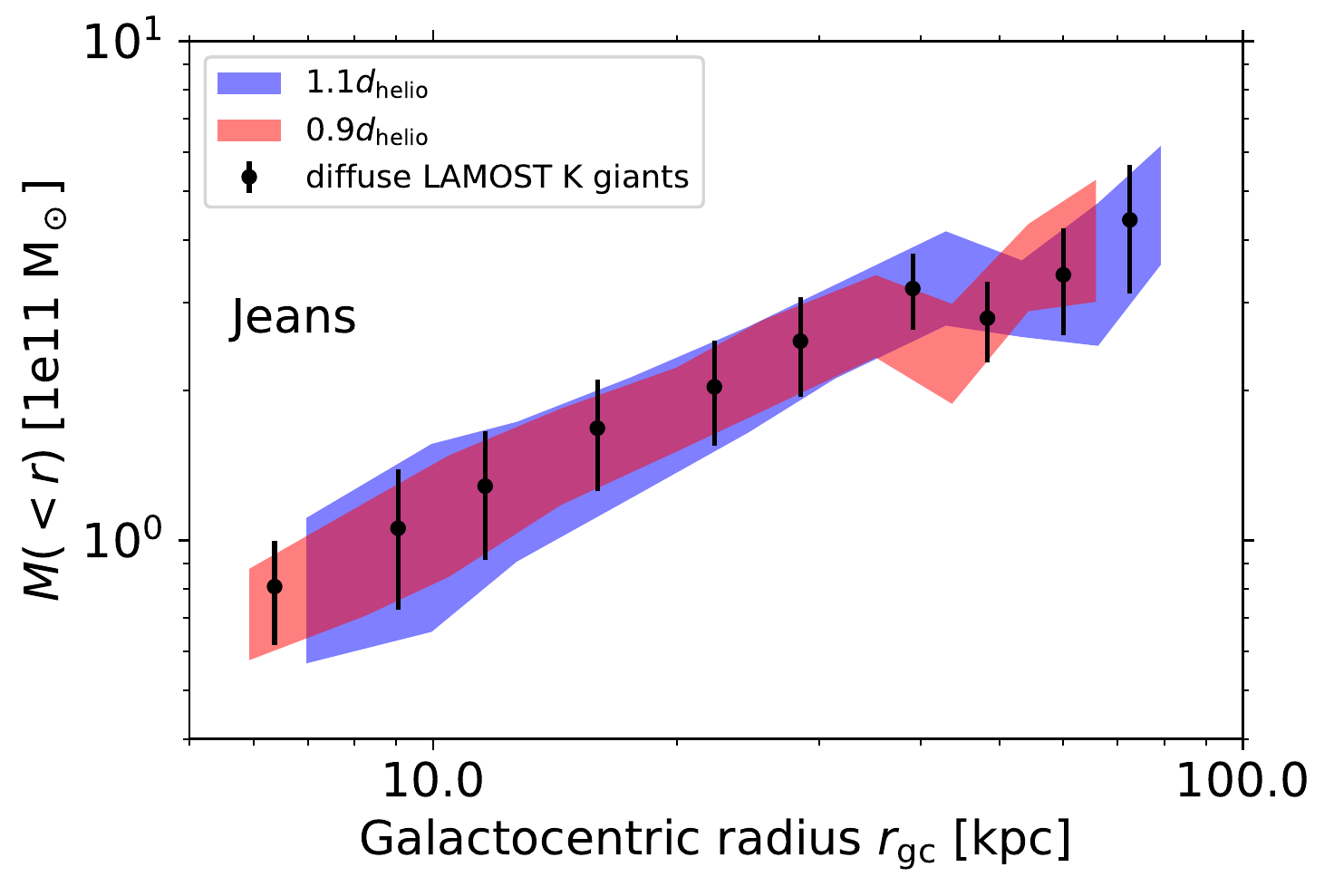}&
\includegraphics[width=.5\columnwidth]{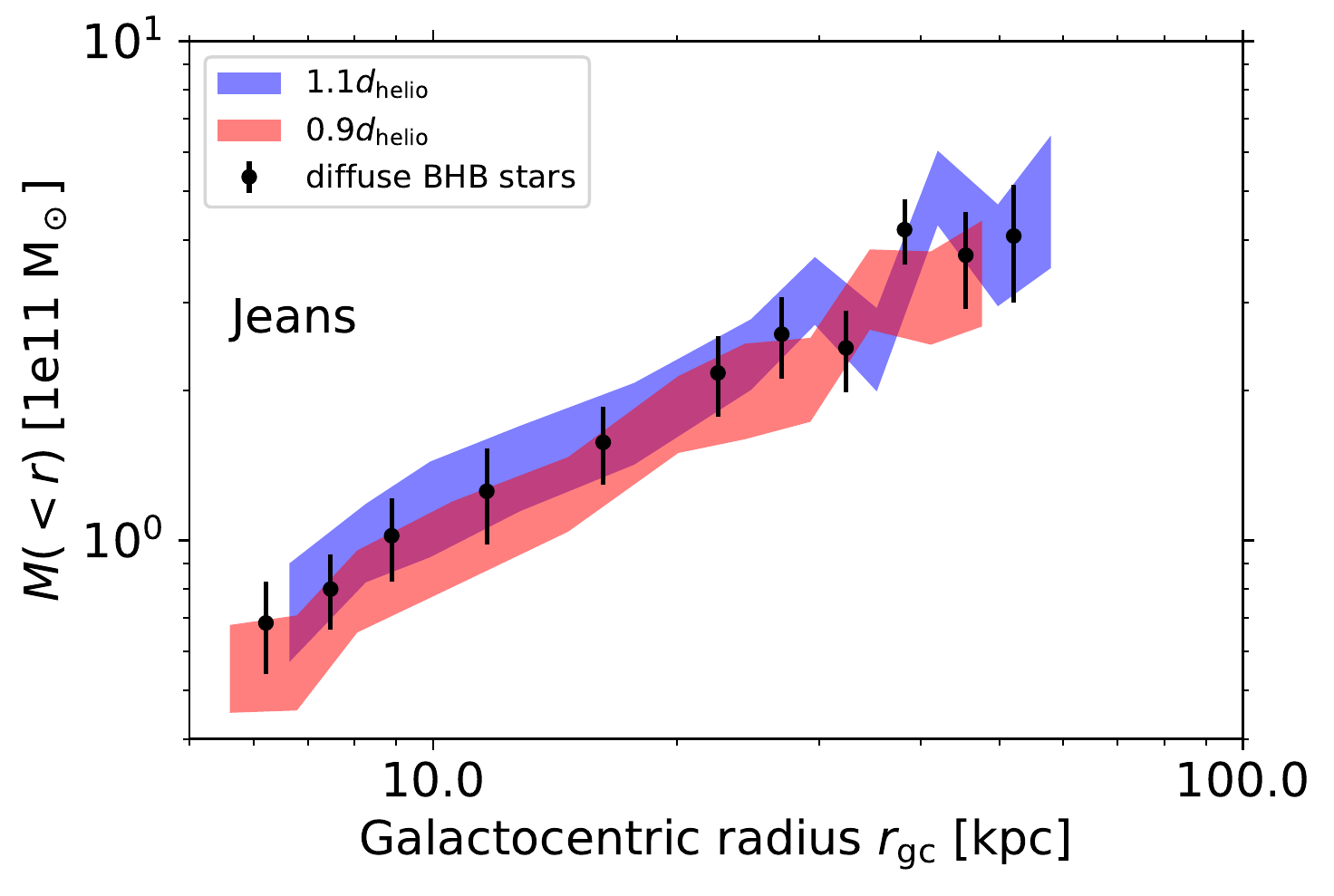}\\
\includegraphics[width=.5\columnwidth]{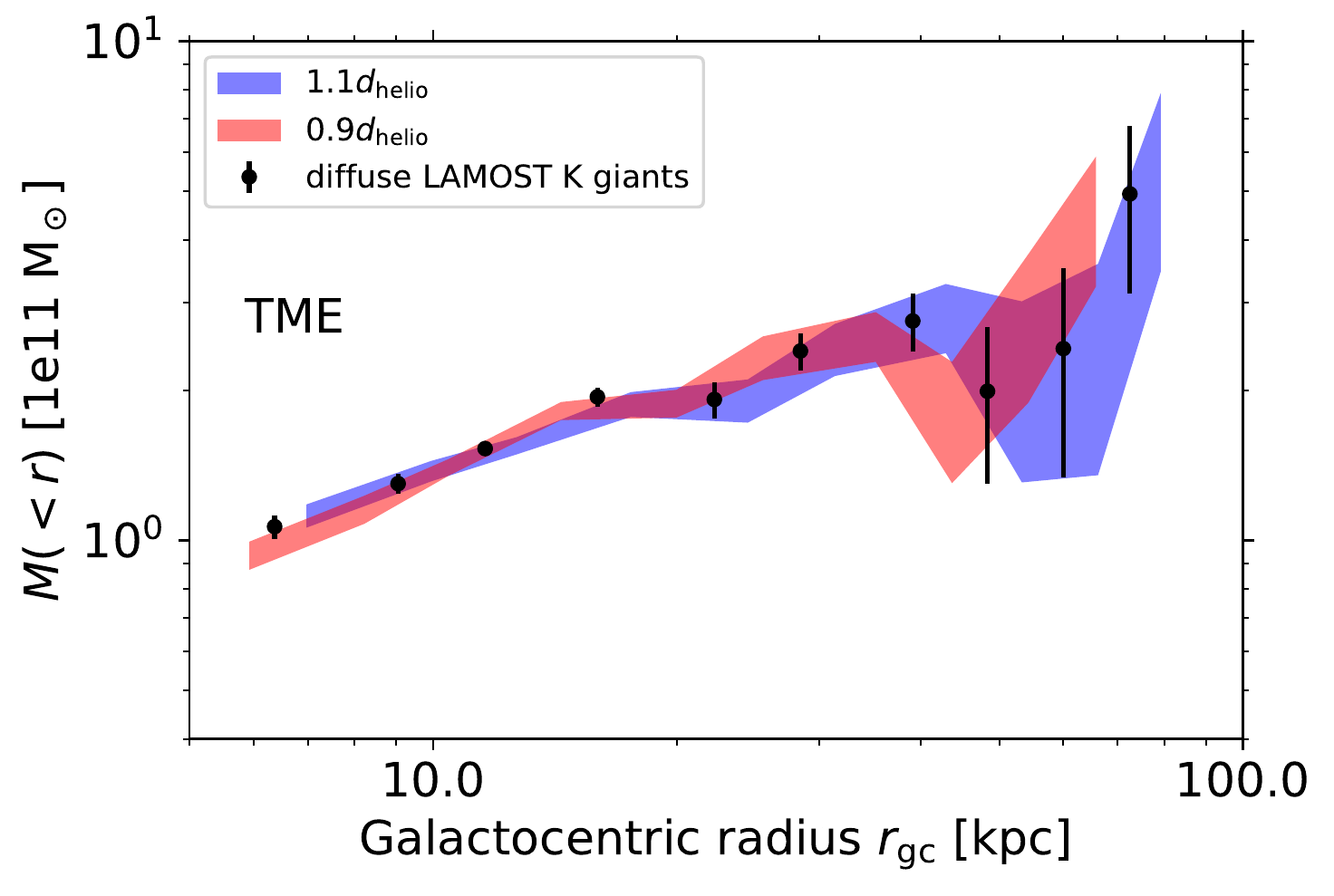}&
\includegraphics[width=.5\columnwidth]{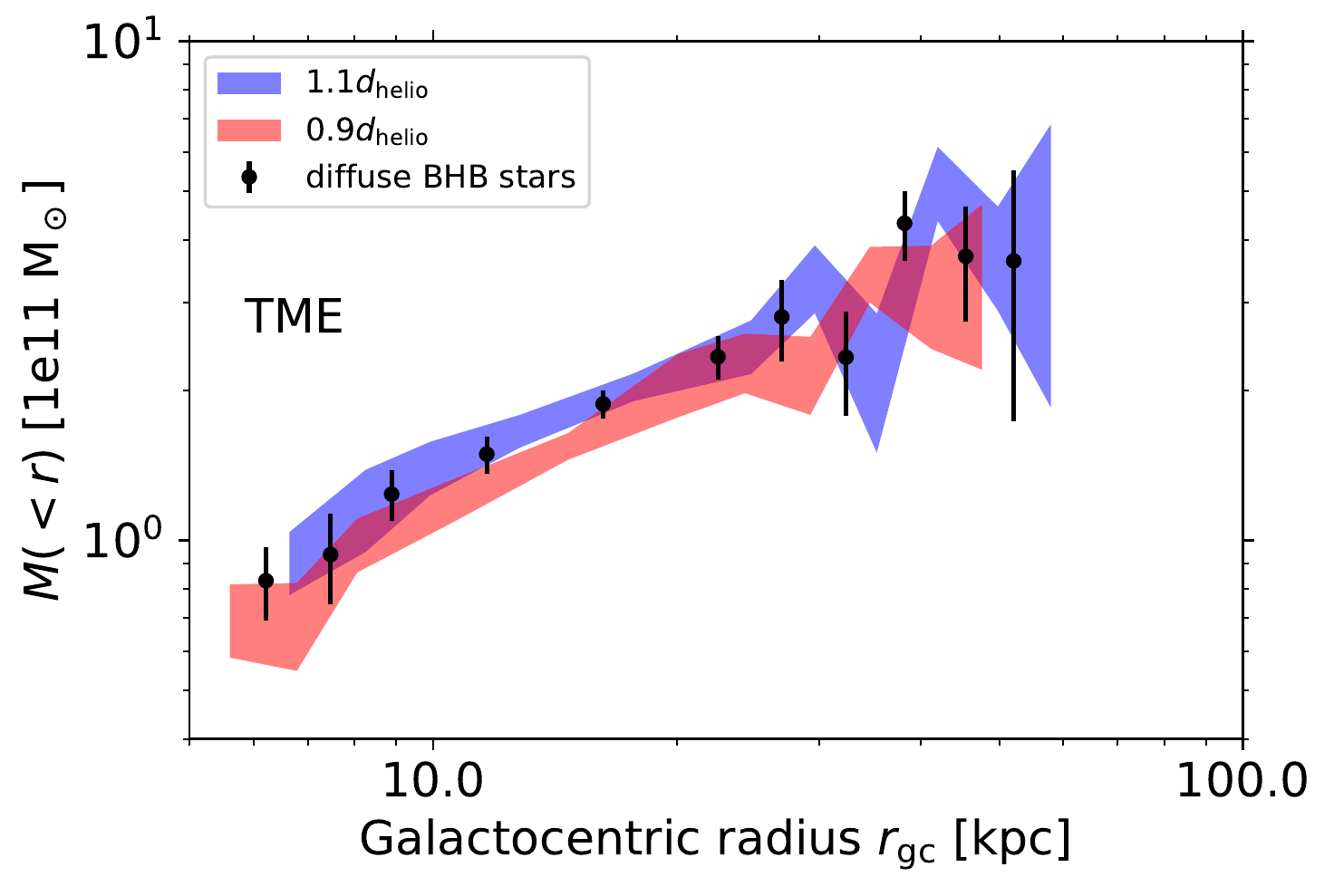}
\end{tabular}
\caption{
Test of distance uncertainties on the enclosed mass $M(<r)$ profile.
We systematically change our distance scales
    for halo LAMOST K giants (left panels) and SDSS BHB stars (right panels) by $\pm10$ percent and remeasure the mass using 3D spherical Jeans (upper panels) and TME (lower panels).
Black symbols show the mass profile for the smooth, diffuse halo LAMOST K giants and SDSS BHB stars using the original distances used for the main results in this work.
Shaded regions show the results of the test for overestimated mass (blue, due to overestimating distances by $1.1d_\mathrm{helio}$) and underestimated mass (red, due to underestimating distances by $0.9d_\mathrm{helio}$). The width of the shaded regions are based on the estimated uncertainties in mass.
Each marker represents the median radius of the stars within our selected radial bins.
}
\label{fig:dist}
\end{figure}

In \citet{Bird2019beta} and \citet{Bird2021} we have
investigated the accuracy and effects due to uncertainty in heliocentric distances $d_\mathrm{helio}$ for KG and BHB stars. We here test the effects of a $\pm10$ percent systematic distance uncertainty for both our KG and BHB stars. This is a generous budget in uncertainty to test. In \citet{Bird2019beta} we detected an approximate 10 percent underestimation for
    our LAMOST halo KG distance scale when compared to that of
    \citet{Bailer-Jones2018} for halo KG stars within 4 kpc of the Sun. We prefer to use the \citet{Xue2014} distance method since it uses globular cluster fiducials observed by SDSS for calibration whereas \citet{Bailer-Jones2018} uses open clusters. The more metal poor globular cluster fiducials are more appropriate for our stellar halo sample than more metal rich open clusters.
\citet{Yang2019b}, using the same KG and BHB sample as our study, select stars with high quality parallaxes within 4 kpc and compare the KG and BHB stars with inverted parallax. Although they find the \citet{Xue2014} KG distances are closer on average by $\sim15$ percent, this bias decreases with fainter {\it Gaia} $G$ magnitudes; thus our preference remains to use the \citet{Xue2014} method as the large bulk of our halo KG stars have fainter $G$ magnitudes. \citet{Yang2019b} find no bias in BHB distances. Although we explore the effects on mass estimation due to biased BHB distances, this is a generous test as we are more confident in the accuracy of BHB distances.

As in Section \ref{sec:density-prof},
we compute the density profiles for spherical shells along
Galactocentric radius $r$ using the \citet{Xu2018} and \citet{Das2016II}
models for our KG and BHB samples, respectively.
The $\pm10$ percent distance uncertainties introduce variations of a few tenths in the power law (both inner and outer) and $\pm1$ kpc in break radius. When needed, we adjust the power law to ensure $200<v_\mathrm{circ}(10\,\mathrm{kpc})<250$ km s$^{-1}$.

In \citet{Bird2019beta} and \citet{Bird2021} we have shown that a $\pm10$ percent distance bias induces changes in the velocity anisotropy $\beta$ of less than one or two tenths.

The changes due to $\pm10$ percent systematic distance uncertainties are shown in Figure \ref{fig:dist} where we present the test results for the Jeans and TME enclosed mass profiles for the LAMOST KG and SDSS BHB smooth halo samples. The changes induced are of order 20 percent.

We estimate the virial mass from these tests and find differences of order 50 percent.


\end{document}